\documentclass[fleqn,usenatbib]{mnras}

\usepackage{newtxtext,newtxmath}
\usepackage[flushleft]{threeparttable}
 \newcommand{\angstrom}{\textup{\AA}}
\usepackage[T1]{fontenc}

\DeclareRobustCommand{\VAN}[3]{#2}
\let\VANthebibliography\thebibliography
\def\thebibliography{\DeclareRobustCommand{\VAN}[3]{##3}\VANthebibliography}

\usepackage{graphicx}	
\usepackage{amsmath}	
\usepackage{stackengine}
\usepackage{threeparttable}

\def\SPSB#1#2{\rlap{\textsuperscript{\textcolor{black}{#1}}}\SB{#2}}

\def\SB#1{\textsubscript{\textcolor{black}{#1}}}
\usepackage[normalem]{ulem}

\newcommand{\ergs}{\mbox{ erg s}^{-1}}
   
\newcommand{\cc}{\mbox{ cm}^{-3}}

\newcommand{\kpc}{\mbox{~kpc}}

\title[Ionization from radio-loud AGNs]{The extent of ionization in simulations of radio-loud AGNs impacting kpc gas discs}

\author[Meenakshi et. al]{Moun Meenakshi$^{1}$,\thanks{E-mail:mounmeenakshi@iucaa.in}
Dipanjan Mukherjee$^{1}$,\thanks{E-mail:dipanjan@iucaa.in} Alexander Y. Wagner$^{2}$, Nicole P. H. Nesvadba$^{3}$,\newauthor Raffaella Morganti$^{4,5}$, Reinier M. J. Janssen$^{6,7}$, Geoffrey V. Bicknell$^{8}$\\
$^{1}$Inter-University Centre for Astronomy and Astrophysics, Pune- 411007, India\\
$^{2}$University of Tsukuba, Center for Computational Sciences, Tennodai 1-1-1, 305-0006, Tsukuba, Ibaraki, Japan\\
$^3$Université de la Côte d'Azur, Observatoire de la Côte d'Azur, CNRS, Laboratoire Lagrange, Bd de l'Observatoire, CS 34229,06304 Nice cedex 4, France\\
$^4$ASTRON, the Netherlands Institute for Radio Astronomy, Oude Hoogeveensedijk 4, 7991 PD, Dwingeloo, The Netherlands\\
$^5$Kapteyn Astronomical Institute, University of Groningen, Postbus 800, 9700 AV Groningen, The Netherlands\\
$^6$Jet Propulsion Laboratory, California Institute of Technology, 4800 Oak Grove Dr., Pasadena, CA 91109, USA\\
$^7$Department of Astronomy, California Institute of Technology, 1216 E California Blvd., Pasadena, CA 91125, USA\\
$^8$Research School of Astronomy and Astrophysics, The Australian National University, Canberra, ACT 2611, Australia
}

\date{Accepted XXX. Received YYY; in original form ZZZ}

\pubyear{2021}

\begin{document}
\label{firstpage}
\pagerange{\pageref{firstpage}--\pageref{lastpage}}
\maketitle

\begin{abstract}

We use the results of relativistic hydrodynamic simulations of jet-ISM interactions in a galaxy with a radio-loud AGN to quantify the extent of ionization in the central few kpcs of the gaseous galactic disc. We perform post-process radiative transfer of AGN radiation through the simulated gaseous jet-perturbed disc to estimate the extent of photo-ionization by the AGN with an incident luminosity of $10^{45}\ergs$. We also map the gas that is collisionally ionized due to shocks driven by the jet. The analysis was carried out for simulations with similar jet power ($10^{45}\ergs$) but different jet orientations with respect to the gas disc. We find that the shocks from the jets can ionize a significant fraction (up to 33$\%$) of dense gas ($n>100\,\mathrm{cm^{-3}}$) in the disc, and that the jets clear out the central regions of gas for AGN radiation to penetrate to larger distances in the disc. Jets inclined towards the disc plane couple more strongly with the ISM and ionize a larger fraction of gas in the disc as compared to the vertical jet. However, similar to previous studies, we find that the AGN radiation is quickly absorbed by the outer layers of dense clouds in the disc, and is not able to substantially ionize the disc on a global scale. Thus, compared to jet-ISM interactions, we expect that photo-ionization by the AGN radiation only weakly affects the star-formation activity in the central regions of the galactic disc ($\lesssim 1$ kpc), although the jet-induced shocks can spread farther out.
\end{abstract}




\section{Introduction}

It is now well established that feedback from Active Galactic Nuclei (AGN) is a crucial ingredient in regulating the evolution of galaxies \citep{silk98,granato_2004,springel_2005,croton_2006,hopkins_2010}. The role of AGN feedback has been broadly studied in the context of outflows, which may either originate from radiation pressure driven winds \citep{zubavous_2012,tombesi15,Vayner_2021} or large scale mechanical jets \citep{Sutherland07,nesvadba11,wagner12,dipanjan16}. Such outflows are expected to regulate the star formation of the galaxy by gas removal \citep{springel_2005,wagner12,hopkins16}, inducing local turbulence \citep{nesvadba10,dipanjan2018,mandal_2021} or by heating the circum-galactic environment to prevent catastrophic cooling flows \citep{dekel_2006,fabian12,gaspari_2011,gaspari_2012}.

However, another mechanism by which the AGN can affect the host's ISM  is by ionizing the cold, potentially star forming gas, either through photo-ionizing radiation from the central nucleus or by collisional ionization from shocks. How deeply the AGN radiation or shocks from the outflows can penetrate the gas in the galaxy is an open question. Several observational studies have emphasized the significance of both AGN radiation and jet-cloud interactions in ionizing the ISM \citep{best_2000,tadhunter_2000,tadhunter_2002,whittle_2002}.
Deep optical spectroscopic observations of several compact \citep{holt_2009,santoro_2020} and large radio sources \citep{solorzano_2002} show that the AGN radiation dominates the ionization in the central regions, while compelling evidence of shock-ionization due to radio jets in extended emission line region (EELR) are also observed in several systems, such as, 3C~171, 3C~277.3, 3C~265 \citep{clark_1998,solorzano_2002,solorzano_2003}, PKS2250-41 \citep{montse_1999}, and 3C~303.1 \citep{shih_2013}. Simulations demonstrate that the jets disperse the gas in the ISM, and induce shock-heating on large scales \citep{Sutherland07,wagner12,dipanjan16}. However, the spatial extent to which radiation and jets ionize the ISM has not been well explored. 

The impact of the photo-ionizing radiation on kpc scales in the host galaxy has been investigated in a few theoretical studies \citep{roos,bieri17}. In \citet{roos}, the simulations from \citet{gabor13} were used in post-process to study the effect of the AGN radiation on the large-scale galactic disc. However, the effect of shock ionization in the ISM was not studied in detail. 
 \citet{bieri17} studied the in-situ effect of AGN ionization in a disc-galaxy using \textsc{Ramses-RT} \citep{rosdahl13}. 
Both of these  papers found that the ionizing AGN radiation fails to penetrate deep into the dense gas. 
The above studies have focused on the impact of non-relativistic radiation-pressure driven winds from the AGN, which are not believed to significantly affect the gas distribution in the host galaxy \citep{gabor13,cielo_2018}. 

Similar to the winds, the relativistic jets can have a substantial impact on the local ISM \citep{cielo_2018}. 
Simulations of radio jets emerging out of the host's ISM show that the jets vigorously distort the gas distribution in the gaseous galactic discs \citep{gaibler12,cielo_2018,dipanjan5063,dipanjan2018}. The jet-ISM coupling significantly depends on the jet's inclination with respect to the galactic disc \citep{dipanjan2018}. Thus, jets can also strongly affect the extent of photo-ionization in the disc. Therefore, in this paper, we use the simulations from \citet{dipanjan2018} to examine the effect of photo-ionization due to the central AGN. These simulations studied the jet-ISM interactions for gas-rich discs of radius $\sim 2$ kpc, with jets launched at different inclination angles.
Hence, this work focus on the inner few-kpcs of the disc, where many studies have found evidence of the impact of the jet, while the gas also shows signature of high excitation originating from photo-ionization from the AGN \citep{tadhunter_2002,santoro_2018}. This analysis is performed using photo-ionization and radiative transfer (RT) code \textsc{Cloudy} \citep{cloudy17} applied in post-process to the simulations.

The paper is structured as follows. In Sec.~\ref{method_RT}, we describe our methodology for the RT study using \textsc{Cloudy} and present its results in Sec.~\ref{Results_RT}. We discuss the implications of our main results in Sec.~\ref{discussion} and summarise the findings from our study in Sec.~\ref{summary}.

\section{Methodology for Radiative Transfer study}
\label{method_RT}
In this section, we discuss the simulations used and our method for the radiative transfer (RT) analysis.
The simulations used in this work are taken from \citet{dipanjan2018}, who studied the evolution of relativistic jets  interacting with the turbulent gaseous disc. We use \textsc{Cloudy} \citep[release C17.02]{cloudy17} to perform RT calculations in the galaxy, applied in post-process to these simulations. 
Our analysis is similar to that of \citet{roos}, who looked at the impact of radiation from an AGN in the simulations of non-relativistic AGN winds impacting a galactic disc \citep{gabor13}. In this study, on the other hand, we explore the effects of ionization due to a radio-loud AGN (henceforth RLAGN) that is driving a relativistic jet into the kiloparsec scale gas disk. We explore how different orientations of the jet with respect to the disc affect the ionization of the ISM.

\subsection{Brief summary of the simulation setup}
We give a brief overview of the simulations in this section and refer the reader to \citet{dipanjan2018} for further details. The simulations evolve a pair of relativistic jets propagating through gas-rich disc of radius 2~kpc, using the \textsc{Pluto} relativistic hydrodynamics code \citep{Mignone07}. They are performed in  Cartesian (X-Y-Z) coordinates, with a domain of physical dimensions 4~kpc $\times$ 4~kpc $\times$ 8~kpc, and a grid of size 672 $\times$  672 $\times$  784 cells. The resolution is $\sim 6$~pc along the X and Y direction. Along the Z-direction, the resolution is $\sim 6$~pc for $z\pm 1.6\,$kpc and a stretched grid with geometrically increasing cell size beyond this.
A two-phase ISM was initialised with a dense turbulent inhomogeneous gas disk of radius $\sim 2$~kpc, rotating at approximately the Keplerian velocity. The disc is immersed in a tenuous hot isothermal atmosphere which was initially in hydrostatic equilibrium with an external gravitational field ($\Phi_\mathrm{DI}$). The gravitational field is modelled following a spherical double-isothermal potential.
Atomic cooling is included in the simulations using a non-equilibrium radiative cooling function generated by the \textsc{Mappings V} code \citep{sutherland17} for solar abundances. At temperatures beyond $\sim 10^9~$K, the cooling function is extended to include Bremsstrahlung emission, and cooling is turned off at temperatures below $\sim$1000~K.

We study the ionization extents in three such jet simulations from \citet{dipanjan2018}, each with different inclination angles of the jet with respect to the disc normal. A fourth simulation, which henceforth we refer as the `no-jet' (or control) simulation, considers the evolution of the rotating gas disc without a jet. The list of simulations analysed in this work are presented in Table~\ref{tab:sim_table}.

\begin{table}
	\centering
	\caption{Simulations from \citet{dipanjan2018} used in this study}
	\label{tab:sim_table}
	\begin{threeparttable}
	
	\begin{tabular}{|c|l|l|l|l|l|l|l|l|} 
		\hline
		Simulation & Power ($\mathrm{P_j}$) & $n_{w0}$ & ${\theta_\mathrm{J}}^b$ & $\Gamma^c$ & Gas Mass \\
		label & ($\mathrm{erg\,s^{-1}})$ & ($\mathrm{cm^{-3}})^a$ & & & ($10^{9}\,\mathrm{M_\odot}$)\\
		\hline
		no jet & - & 200 & - & - & 5.71 \\
		B & 10$^{45}$ & 200 & $0^\circ$ & 5 & 5.71\\
		D & 10$^{45}$ & 200 & $45^\circ$ & 5 & 5.71\\
		E & 10$^{45}$ & 200 & $70^\circ$ & 5 & 5.71\\
		\hline
	\end{tabular}
	\begin{tablenotes}
	    %
	     \item[a] The density of the dense gas at the center of the disc.
         \item[b] Angle of inclination of the jet measured from the normal to the disc.
       \item[c] The Lorentz factor of the jet.
       \item[] In the jetted simulations, the jet was injected after the disk had been evolved for $t\sim0.17$~Myr, to allow it to settle briefly. In \citet{dipanjan2018}, the jet injection time was re-scaled to $t=0$, and they subsequently discussed the time evolution relative to the injection time. In this work however, we use the original time of the simulation. Hence the simulation times presented in this work and \citet{dipanjan2018} will be offset by $\sim 0.17$~Myr, for the same snap-shot.
	\end{tablenotes}
	\end{threeparttable}
\end{table}

\begin{figure*}
\centerline{
\def\arraystretch{1.0}
\setlength{\tabcolsep}{0.0pt}
\begin{tabular}{lcr}
    \hspace{-1cm}
    \includegraphics[width=0.4\linewidth]{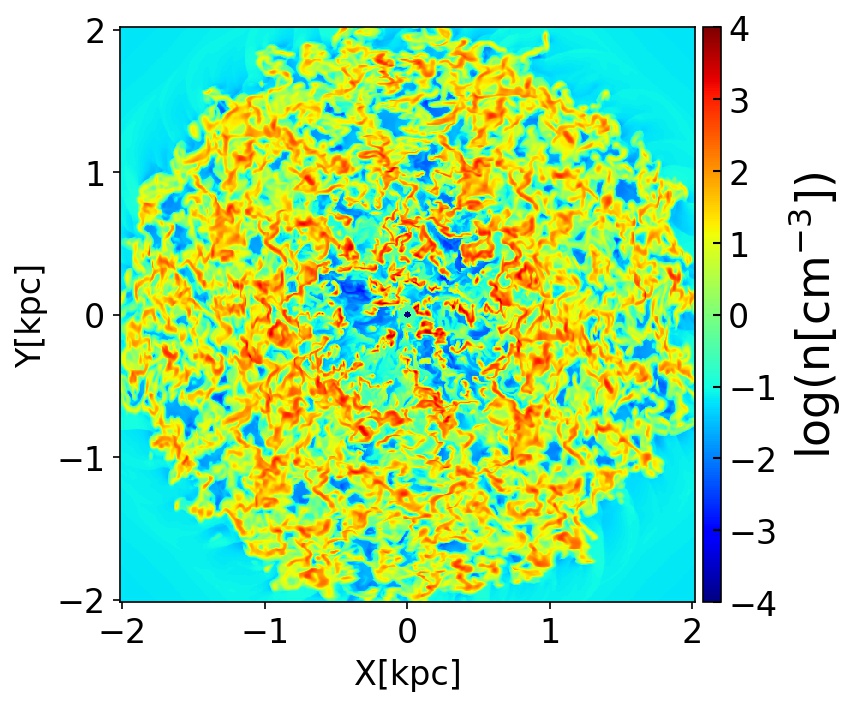} &
    \includegraphics[width=0.4\linewidth]{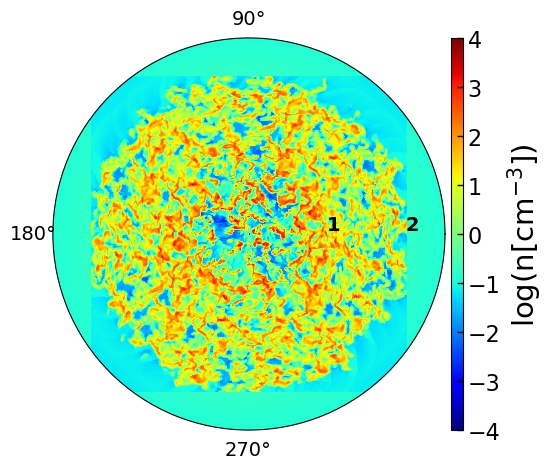}
\end{tabular}}
    \caption{Density $\mathrm{(\log\,n[cm^{-3}])}$ in the disc plane for Sim.~B ($\theta_\mathrm{J}=0^{\circ}$) at 0.68~Myr in the Cartesian grid (Left) and the spherical grid (Right). In the right panel, the radial distance from the center is noted with bold black numbers in units of kpc. The extrapolation in the newly added cells is done using the density profile of the hot halo from \citet{dipanjan2018}, which is shown in eq.~\ref{eq:halo}.}
    \label{fig:mesh}
\end{figure*}

\subsection{Computations using \textsc{Cloudy}}
\label{cloudy_comp}
The photoionization and radiative transfer code {\textsc{Cloudy}} is capable of calculating the ionization and thermal equilibrium state of gas exposed to an external radiation field along a one-dimensional column of gas. This is achieved by solving the equations of statistical and thermal equilibrium simultaneously, which balances ionization and recombination of different elements, and the resulting heating and cooling of the gas. These calculations are performed cell-by-cell along the column of gas, and are time-independent. 
This analysis is purely in post-process, and any thermodynamic changes calculated with \textsc{Cloudy} are assumed to be sufficiently small that they would have a negligible effect on the gas dynamics in the simulations compared to the effect of the energy transfer to the gas by the jet. We verify the validity of this assumption a posteriori and confirm that the momentum transfer from the AGN radiation to the gas is almost negligible as compared to the jet, as shown in Appendix~\ref{mom-compare}. This assumption breaks down if the AGN radiation is sufficiently powerful (see Appendix~\ref{mom-compare}) and the gas is sufficiently dust-rich for infrared photons to become important in driving outflows \citep{bieri17, cielo_2018}. A full investigation of the interplay of radiation and jet-driving would require a hydrodynamic simulation treating both simultaneously. Our analysis here is performed in the limit of negligible thermal and dynamical back-reaction due to radiation.

The radiative transfer with \textsc{Cloudy} is performed along one-dimensional radial Lines of Propagation (LOP), outward from the central AGN.  To perform the \textsc{Cloudy} computations, the grid from the hydrodynamic simulations is transformed to a spherical domain ($r$, $\theta$ and $\phi$) with dimensions: 4.9~kpc $\times$ $180^\circ\,\times$ 360$^\circ$,  represented by a grid of size $816\times524\times1048$ cells. The polar angle $\theta$ is measured from the positive Z-axis, and the azimuthal angle $\phi$ is measured from the positive X-axis. The resolution for the spherical mesh is 6~pc along the radial direction and 0.006 rad for both the $\theta$ and $\phi$ coordinates.
This gives a total of 549152 (product of the number of cells along $\theta$ and $\phi$) Lines of Propagation (LOPs) to analyze the ionization structure in the disc. The grid has been chosen such that the central region of the galaxy at a radius of $\sim 1$~kpc, is resolved with a volume element of $216\, \mathrm{pc}^3$, similar to that of the original Cartesian grid. The hydrodynamic variables from the simulations, such as density and pressure, are then mapped from the rectilinear hydrodynamic simulation grid to the spherical grid using tri-linear interpolation.

The two panels in Fig.~\ref{fig:mesh} show the disc plane (X-Y plane with $\theta=90^\circ$) of Sim.~B at 0.68~Myr in the Cartesian and spherical grids respectively. For cells in the spherical grid lying beyond the domain of the original Cartesian grid, the density and temperature values were extrapolated from the expected values of the hydrostatic gas in the halo \citep{dipanjan2018}:
\begin{equation}\label{eq:halo}
    n(r)=n_{\mathrm{h0}}\,\mathrm{exp}\left[\frac{-\Phi_\mathrm{DI}(r)}{k_BT/(\mu m_a)}\right]\, ,\,\, T=10^7 K.
\end{equation}
Here $n_{\mathrm{h0}}$ is the central density at $r$=0, where $r$ is the spherical radius, $\mu$ is the mean molecular weight ($\mu\sim0.6$), $m_a=1.6605\times 10^{-24}\,\mathrm{g\,cm^{-3}}$ is the atomic mass unit, $k_B$ is Boltzmann's constant, and $T$ denotes the temperature of the gas in the halo. Since the halo region contains low density ($<1\,\mathrm{cm^{-3}}$) ionized gas, the extrapolation of the density profile in the spherical grid has a negligible effect on the calculations of AGN ionization, as discussed in Sec.~\ref{shocks}.

The RT computation along an LOP terminates either when the equilibrium electron kinetic temperature in a zone drops below 4000~K, when a characteristic ``ionization radius'' is reached, or when the end of the LOP is reached. We define the ionization radius along an LOP to be the distance to the first cell in the LOP in which at most 50$\%$ of the hydrogen gas mass in the cell is ionized, and also use it as the stopping criteria for the RT. The ionization radius, thus, is a measure of the distance to which the AGN radiation has significantly ionized the medium. 

The RT calculations with \textsc{Cloudy} along an LOP can be run iteratively, where for better accuracy, subsequent iterations dynamically refine the grid at regions of high gradients in the optical depth. However, we have verified that additional iterations do not affect our analysis: they change the value of the mean ionization radius in the disc plane ($\theta=90^{\circ}$) by less than 1$\%$. Hence, we only perform one RT calculation per LOP, to avoid the computational cost of additional iterations.

\begin{figure*}
 \centerline{
\def\arraystretch{1.0}
\setlength{\tabcolsep}{0.0pt}
\begin{tabular}{lcr}        
    \hspace{-1cm}
    \includegraphics[width=0.5\linewidth]{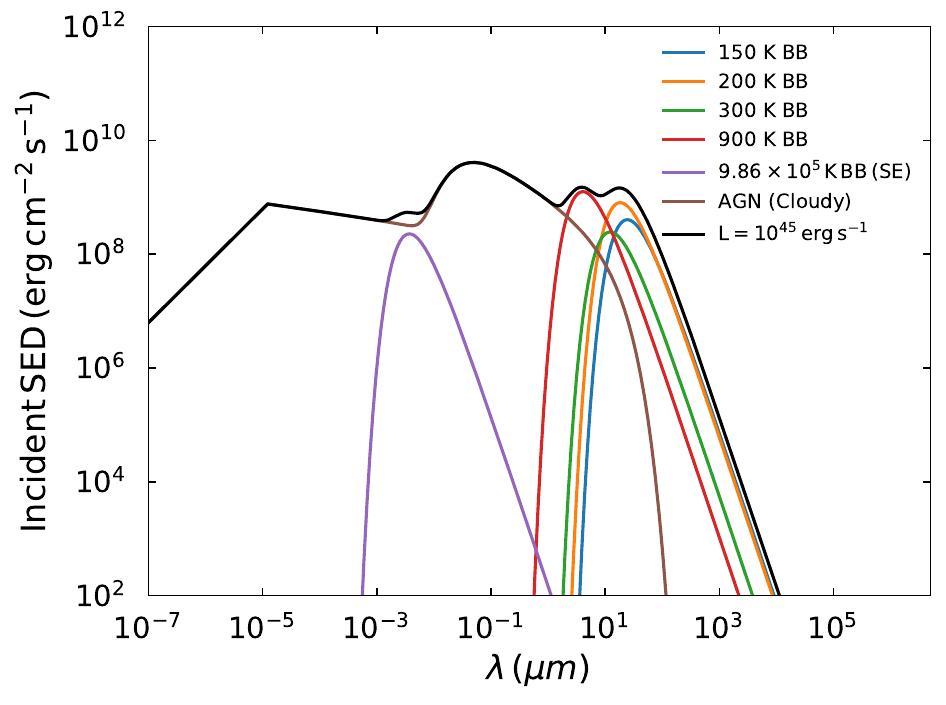} &
    \includegraphics[width=0.5\linewidth]{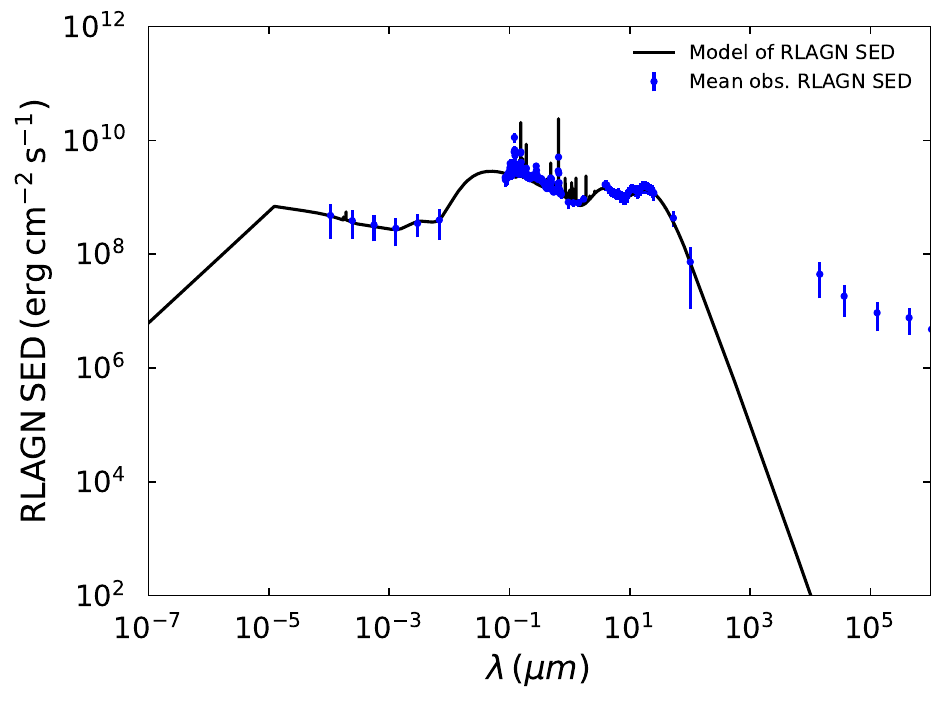}    \end{tabular}}
    \caption{\textbf{Left:} Incident SED of the inner region of a Radio Loud AGN of luminosity $10^{45}\,\mathrm{erg\,s^{-1}}$, which is the
the sum of the emission from an AGN and multiple black-body (labelled BB) components from a dusty torus. The component labelled `SE' refers to the soft-excess emission, modelled by a black-body.
\textbf{Right:} Emergent SED of the Radio loud AGN, i.e., (0.7$\times$ Incident SED +0.3 $\times$ Transmitted through the Broad Line Region (BLR), fitted to the mean SED from \citet{shang}}
 \label{fig:sed}
\end{figure*}

\subsection{Spectral Energy Distribution (SED) of the AGN}
\label{SED}

We construct an AGN SED considering contributions from various emission components following the AGN unification model. The different elements of the AGN SED are chosen in a way such that the combined SED closely matches the mean AGN spectrum from \citet{shang}. The different components of the modelled SED are as follows:
\begin{itemize}
    \item \textbf{Nuclear AGN emission:\,}
The multi-component nuclear AGN spectrum taking into account both the high-energy power law and the UV big blue bump is described by the following equation \citep{korista_1997,roos}:
\begin{equation}
     F_{\nu}=\nu^{\alpha_{\mathrm{UV}}}e^{-\left(\frac{h\nu}{k_B T_{\mathrm{BB}}}\right)}e^{-\left(\frac{k_B T_{\mathrm{IR}}}{h\nu}\right)}+a\nu^{\alpha_\mathrm{X}}.
 \end{equation}
 Here $T_{\mathrm{BB}}$ is the cut-off temperature of the blue bump, $T_{\mathrm{IR}}$ is the IR cut-off temperature, and $\alpha_{\mathrm{UV}}$ and $\alpha_{\mathrm{X}}$ are the exponents in the UV and in the X-ray fields, respectively. The coefficient $a$ is adjusted so that the ratio of the emission between optical and X-ray bands are given by a pre-specified index $\alpha_{OX}$ as:
 \begin{equation*}
     \frac{F_\nu(2\mathrm{keV})}{F_\nu(2500\angstrom)}=\left(\frac{\nu_\mathrm{2keV}}{\nu_{2500\angstrom}}\right)^{\alpha_{\mathrm{OX}}}=403.3^{\alpha_{\mathrm{OX}}}
 \end{equation*}
 The AGN spectrum was evaluated using \textit{AGN} command in \textsc{Cloudy} with the following SED parameters: $T_{\mathrm{BB}}=\mathrm{4\times10^5}$~K, $\alpha_{\mathrm{UV}}=-0.3$, $\alpha_{\mathrm{X}}=-0.85$ and $\alpha_{\mathrm{OX}}=-1.28$. The parameters used above gives an SED that matches well with the spectrum in \citet{shang}, and are similar in values used in \citet{elvis} and \citet{roos}.
 \item \textbf{Black-body emission from a heated dusty torus:\,} Infra-red (hereafter IR) emission from a dusty torus was modelled using multiple black bodies with temperatures ranging from  150~K to 900~K. The infrared radiation does not contribute to the ionization of the ISM, and hence the RT calculations do not depend on this part of the spectrum. However, the IR component has been introduced for completeness and correct evaluation of the total bolometric luminosity.
\item \textbf{Soft-excess emission:\,} The soft excess is a crucial component in the spectra of many AGN in the X-ray band. The soft-excess emission is believed to originate from either Comptonization of soft X-ray (EUV) disc photons in a hot plasma \citep{porquet04}, or reflection from a photoionized accretion disc around a black hole \citep{ross93,crummy06}. These high-energy photons can contribute to ionizing the gas in the galaxy.
The soft-excess component in various sources is fitted with black-bodies with temperatures ranging in 85-160 eV \citep{crummy06,laha}. In our study, we model soft-excess X-ray emission in the SED by defining a black body with an effective temperature of 9.86$\,\times\,10^5 $~K ($k_BT \sim$ 85 eV), as done in \citet{laha}.

We set the total bolometric luminosity of the incident spectrum, which includes the above three components (nuclear AGN spectrum, soft-excess emission and the IR emitting black-bodies), to $10^{45}~\ergs$, similar to the kinetic power of the relativistic jets injected into the simulation. We use this incident SED to compute the transmitted spectrum reprocessed by the Broad Line Region (BLR), as discussed below.
\item \textbf{Broad Line Region (BLR):\,} The transmitted BLR spectrum is obtained after propagating the incident AGN Spectrum (see left panel in Fig.~\ref{fig:sed}) through homogeneous clumps with a gas density of $10^9\,\mathrm{cm^{-3}}$ and a filling factor of $10^{-3}$ \citep{gary}, representing a typical BLR. The gas distribution, kinematics, and extent of this region are still not well understood and are believed to have a small covering factor ($<0.3$) \citep{carswell,netzer93,peterson}.
We use a mean covering fraction of 0.3 for the BLR, as suggested by photo-ionization models \citep{netzer93}, to compute the final AGN SED. 
\end{itemize}
 We obtain a transmitted luminosity of 4.39 $\times 10^{44}\ergs$ after processing the incident spectrum from the BLR. We then construct the final solid angle averaged SED of the RLAGN by superimposing 70$\%$ of the incident and 30$\%$ of the transmitted spectrum, which is shown in Fig.~\ref{fig:sed}. 
 The \textsc{Cloudy} scripts for the AGN SED are publicly available on the Github repository\footnote{\url{https://github.com/mounmeenakshi/Cloudy_scripts/tree/main}}.
We assume the dust to be distributed according to solar abundance, and the ISM abundance is Milky-Way-like. The grain sublimation\footnote{This includes graphite and silicate grains in the ISM. We refer the reader to \textsc{Cloudy} release C17.02 Hazy documents for further details.} is taken into account while performing the radiative transfer calculations.

\begin{figure}
    \centering
    \includegraphics[scale=0.45]{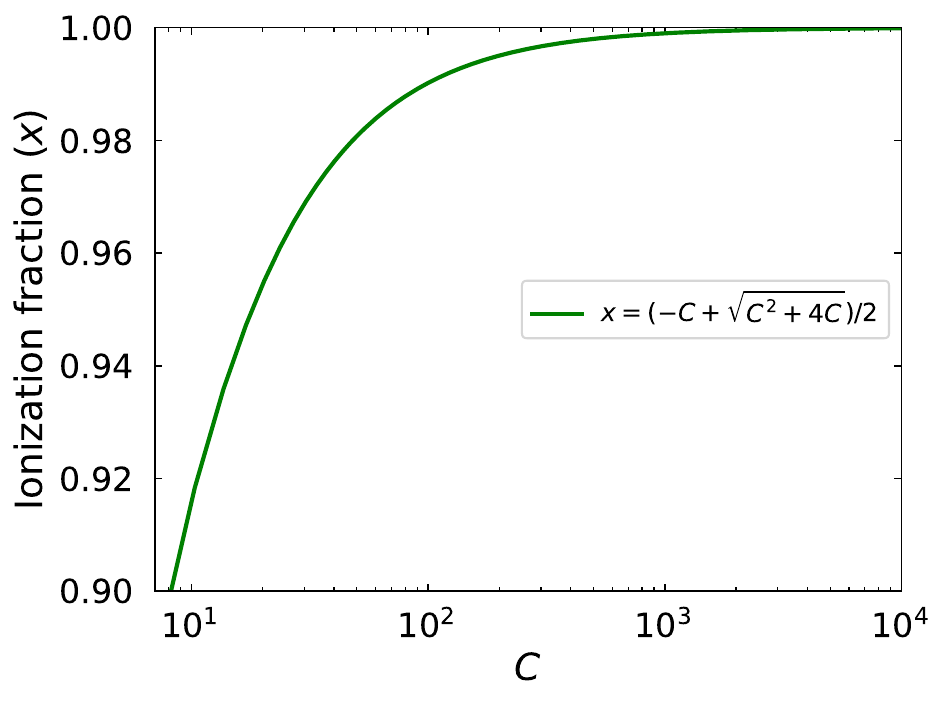}
    \caption{The ionization fraction as a function of R.H.S. term in Saha ionization Equation (denoted by $C$ in eq.~\eqref{saha})}
    \label{fig:saha}
\end{figure}

\begin{figure*}
 \centerline{
\def\arraystretch{1.0}
\setlength{\tabcolsep}{0.0pt}
\begin{tabular}{lccr}
    \includegraphics[width=0.26\linewidth]{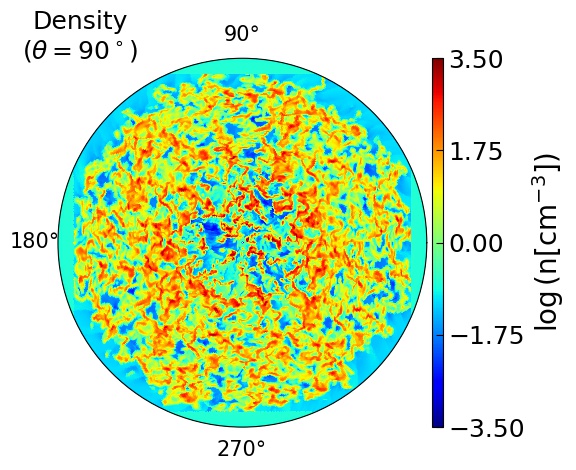} &
    \includegraphics[width=0.23\linewidth]{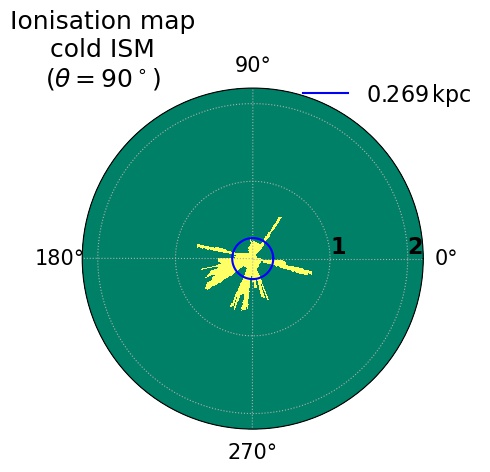} &
    \includegraphics[width=0.26\linewidth]{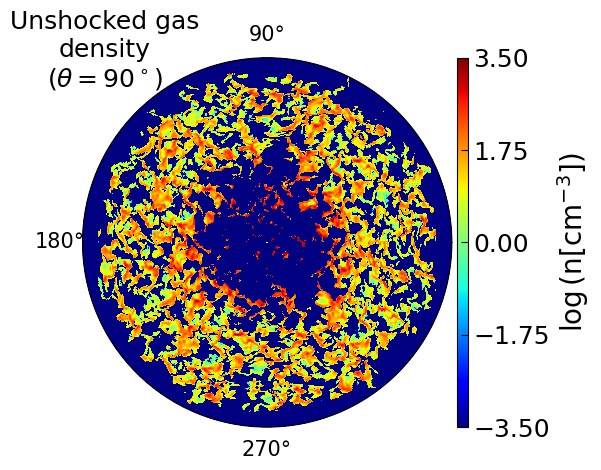} &
     \includegraphics[width=0.23\linewidth]{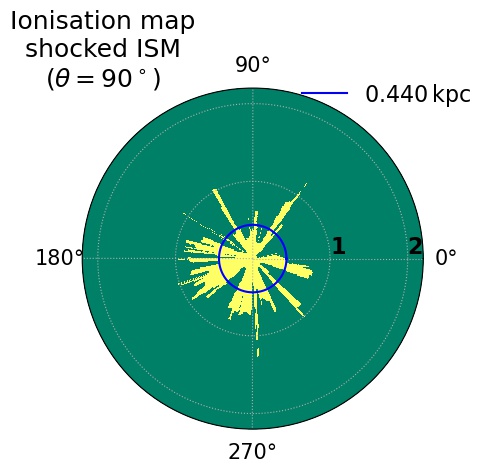} 
     \end{tabular}}
     \caption{Logarithmic number density and extent of ionization (using the AGN SED in Fig.~\ref{fig:sed}) in the mid-plane ($\theta=90^\circ$) of the simulation B at 0.68~Myr. . The ionization maps show the extent of photoionized regions (in yellow). The left two panels show the density and ionization without considering the ionization from shocks. The third panel from the left shows the density after excluding the shocked gas (see Sec.~\ref{shocks} for details) and the last panel shows the corresponding ionization map. The mean ionization radius in the disk plane (presented as $R$ in text) is denoted by the blue circle. The plots show that excluding the shocked gas regions increases the extent of photo-ionized regions in the galaxy.
     }
\label{fig:cold-shock}
\end{figure*}

\subsection{Including ionization due to jet-induced shocks}
\label{shocks}
Shocks induced by the jet \citep{moy,sanderson,dipanjan16,dipanjan2018} can ionize the gas in the central region of the galaxy. The ionized gas will not contribute to the opacity for ionizing photons from the AGN. Thus, assuming the whole gas in the ISM to be neutral underestimates the radiative feedback from the AGN. We thus exclude the shocked regions from the radiative transfer calculations for calculating the extent of photoionization, as explained below.
Assuming that the gas in every cell is in local thermodynamic equilibrium (LTE), we use the Saha ionization equation \citep{paddy2006} to calculate the collisionally ionized gas fraction in a cell:
\begin{equation}{\label{saha}}
    \frac{x^2}{(1-x)}\,=\,\frac{1}{n}\left(\frac{2\pi m_e k_B T}{h^2}\right)^{3/2} e^{(-\frac{\Delta E}{k_BT})} \equiv C
\end{equation}
Here $h$ is the Planck's constant, $m_e$ is the mass of the electron, $k_B$ is the Boltzmann constant, and $\Delta E$ is 13.6 eV, the ionization energy from the ground state in the hydrogen atom\footnote{The collisional ionization estimates in this study assume all the cells to be filled with hydrogen atoms only. But, we verified that assuming Milky-Way-like ISM abundances of major elements i.e., $\mathrm{H}\sim90.8\%,\,\mathrm{and}\, \mathrm{He}\sim9.1\%$ by number density \citep{katia_2001} and including first ionization of helium has no significant effect on the fractions of collisionally-ionized gas mass in Table~\ref{tab:tab3}}.. The ionization fraction is denoted by $x$, and $T$ and $n$ are the temperature and total gas density of the cell, respectively. Fig. \ref{fig:saha} shows that the ionization fraction increases as a function of $C$ and asymptotically reaches 1 for $C\rightarrow \infty$. In our study, we use the value $C\geq1000$, corresponding to $99.9\%$ ionization, to denote regions in which collisional ionization by shocks strongly dominate. Such ``collisionally-ionized'' computational cells are filled with a very low density ($6 \times 10^{-5}\mathrm{cm^{-3}}$), effectively removing their contribution in the radiative transfer calculations of \textsc{Cloudy}. The AGN radiation is thus able to pass through such shocked regions without being absorbed. For lower values of C, the ionization fraction is calculated by solving Eq.~\eqref{saha} to find $x$. 
The density in each cell is then replaced by the non-ionized gas density, given by $(1-x)n$. The results are not sensitive to the limiting value chosen for C, as long as it is much larger than unity.
The RT study of the new density profiles, with the ``collisionally-ionized'' gas cells removed or their density modified, effectively evaluates the ionization due to the AGN radiation of only the unshocked ISM \footnote{For the hydrogen gas density along an LOP that goes into \textsc{Cloudy}, we input the density of unshocked ISM. We confirm that the fractions of photo-ionized gas mass in Table~\ref{tab:tab3} are not substantially affected by neglecting the ionization state of other species in the input in \textsc{Cloudy}.}.

\begin{figure*}
 \centerline{
\def\arraystretch{1.0}
\setlength{\tabcolsep}{0.0pt}
\begin{tabular}{lcr}
    \includegraphics[width=0.33\linewidth]{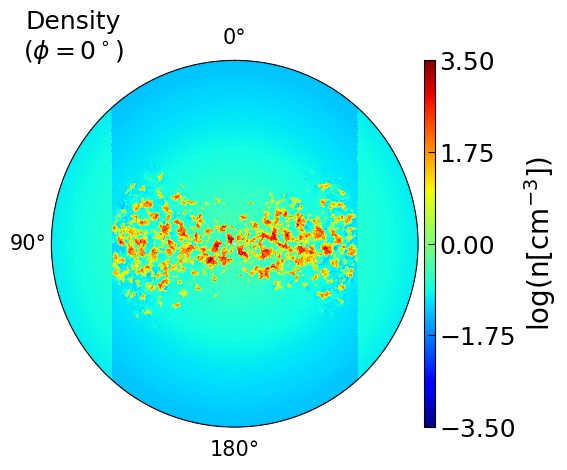} &
     \includegraphics[width=0.33\linewidth]{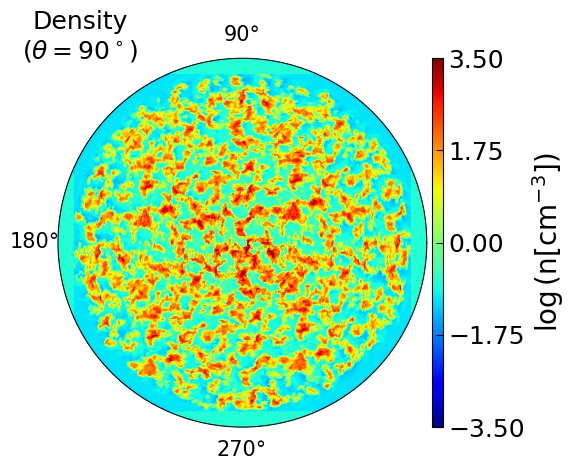} & 
    \includegraphics[width=0.33\linewidth]{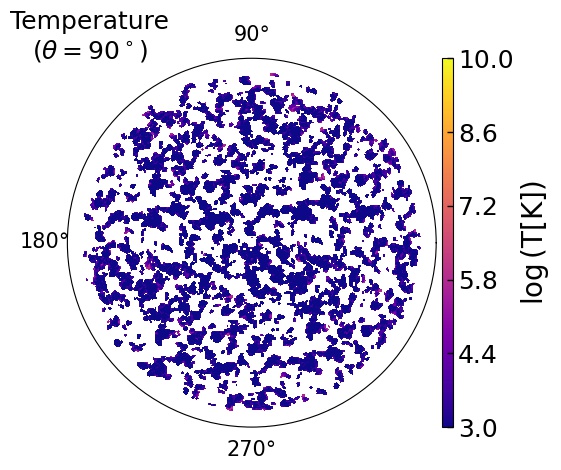}
     \end{tabular}}
\centerline{
\def\arraystretch{1.0}
\setlength{\tabcolsep}{0.0pt}
    \begin{tabular}{lcr}
    \hspace{-0.2cm}
     \includegraphics[width=0.3\linewidth]{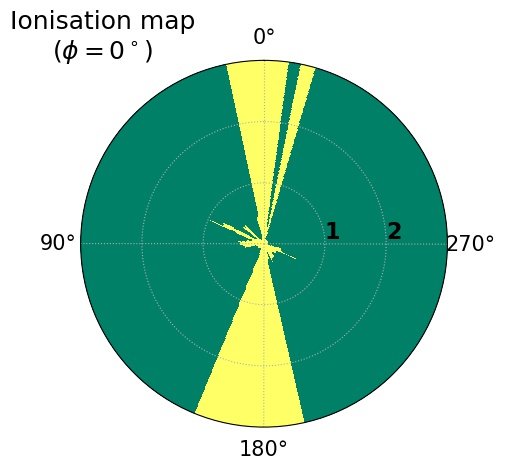} &  \hspace{0.3cm}
    \includegraphics[width=0.31\linewidth]{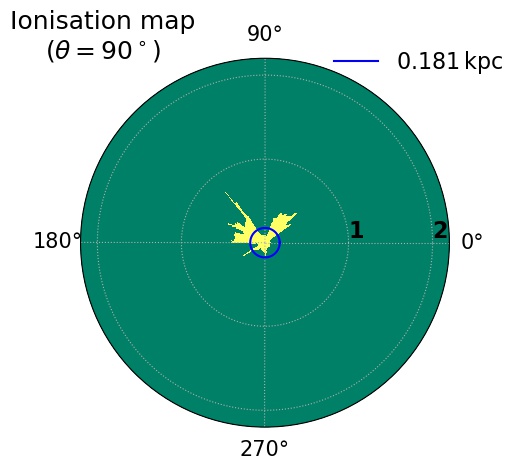} &  \hspace{0.1cm}
    \includegraphics[width=0.35\linewidth]{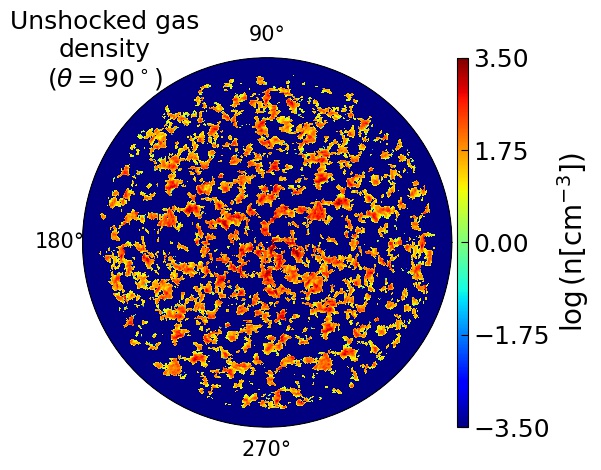} 

     \end{tabular}}
     \caption{Density, temperature and ionization maps for the `no-jet' (or control) simulation at 0.156~Myr. \textbf{First Row:} Left: Density ($\log \mathrm{n[cm^{-3}]}$) slice in vertical ($\phi=0^\circ$) plane (Extrapolating the cells using the initial halo profile (at $t=0$) results in the sharp variations in the density of the vertical plane), Middle: Density in disc ($\theta=90^\circ)$ plane, Right: Temperature ($\log \mathrm{T[K]}$) of the clouds in the disc plane. We use a cloud tracer of 0.99 to eliminate the regions dominated by diffuse ionized gas from the temperature plot. \textbf{Second row:} Left and Middle: Ionization maps in the vertical and disc plane. The blue circle denotes the mean ionization radius in the disc plane (presented as $R$ in text). Right: Density in the disc region after excluding hot gas regions in the settling disc.}
\label{fig:no-jet}
\end{figure*}

 \begin{figure}
  \centering
{\includegraphics[width=0.8\linewidth]{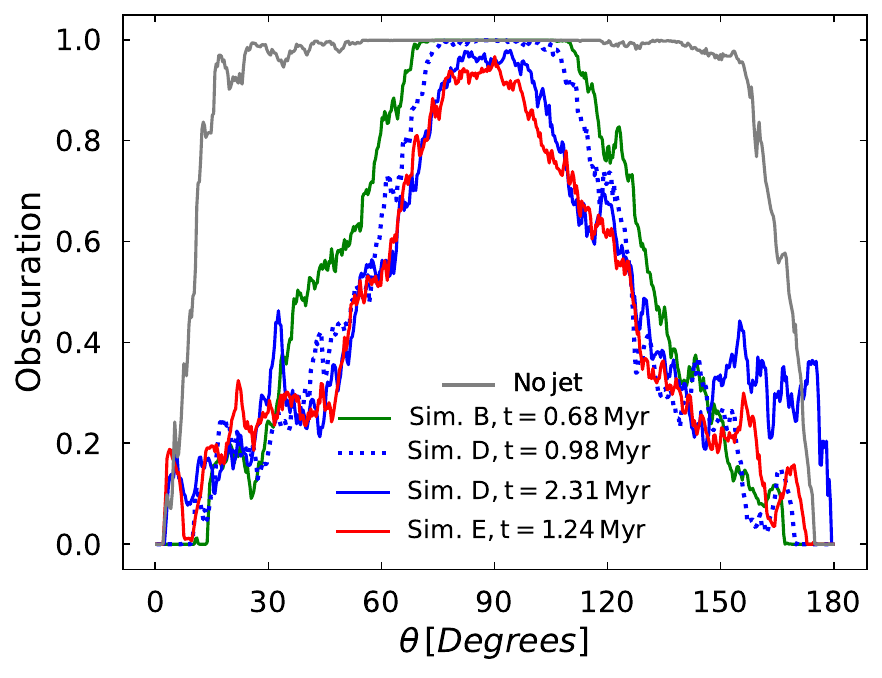}}
\caption{ 
Obscuration fraction as a function of polar angle from the vertical axis, which shows the fraction of regions obscured by neutral gas (e.g., the obscuration is 0 for fully ionized region, and 1 for fully obscured regions). The obscuration is higher closer to the poles in the inclined jets as compared to the case of a vertical jet. Note that the obscuration closer to the poles is increased for Sim.~D ($\mathrm{\theta_J}=45^\circ$) as jet the deflects from its original trajectory and escapes along the vertical axis, i.e. from 0.98~Myr to 2.31~Myr.}
  \label{fig:jet-disc}
\end{figure}

\section{Results}
\label{Results_RT}
We perform the post-process radiative transfer (RT in short) calculations on the simulations presented in \citet{dipanjan2018}. The list of chosen simulations is presented in Table~\ref{tab:sim_table}, where the same naming conventions have been used, as in \citet{dipanjan2018}. 

In the results presented in the subsequent sections, we show the ionization maps in the disc ($\theta=90^\circ$) and vertical planes ($\phi=0^\circ$). We show the mean ionization radius ($R$) in the disc plane (represented as a blue circle)\footnote{$R=\frac{\sum(R_1+R_2+...R_N)}{N}$, where $R_1, R_2,..., R_N$ represents the ionization radius along the LOPs in the disc plane.}. For the fully ionized LOPs we consider the ionization radius to be 2~kpc, equal to the radius of the gas disk. However, the mean ionization radius in the disc plane can not represent the complex statistics of these systems. Therefore in Sec.~\ref{dist_radi}, we discuss the distribution of the ionization radii for different simulations.
To estimate the impact of the ionization on the gas disk, we have also defined some diagnostic quantities, which are shown in Table~\ref{tab:tab3}, and whose values are compared for the different simulations.

\subsection{Ionization of a cold ISM by the AGN and effect of shocks}

In this section, we compare the RT results for the disc plane in Sim.~B (jet perpendicular to the disc) at 0.68~Myr with and without including the effect of shocks. The ISM is assumed to be cold when we do not consider the the effects of shocks, which is similar to the analysis of  \citet{roos}.
The density and the ionization maps of ISM in the disc plane ($\theta=90^\circ$), with and without the shocked gas, are shown in Fig.~\ref{fig:cold-shock}. The ionized regions\footnote{Computational cells in the LOPs ionized to 50$\%$ or more} are represented in yellow in the ionization map. The green regions mark the LOPs (Line Of Propagation) ionized to less than 50$\%$. The third figure shows the disc plane without the shocked gas. The jet injects energy into the central region of the disc and heats it through shocks, which collisionally ionizes the neutral gas. This creates a circular cavity in the center, which is transparent to the AGN radiation (see Sec.~\ref{shocks}). Hence, the mean ionization radius in the disc plane (shown as blue circle) increases to 0.44~kpc, from its former value of 0.269~kpc in the case of the cold ISM. This indicates that the inclusion of effects of shocks from the jet enhances the ionization from the AGN radiation. 

In our study, we find that jet-driven outflows ionize a substantial portion of the gas in the galaxy, which is up to 16$\%$ of the dense mass for the perpendicular jet (Sim.~B) and 33$\%$ for a jet almost parallel to the disc plane (Sim.~E), as shown in Table~\ref{tab:tab3}. This fraction is much higher than the ionized gas mass fraction in the disc without a jet (which is 0.34$\%$).

Thus, assuming the ISM to be cold and neutral in the galaxies underestimates the extent of ionization due to the central engine, as shown in Fig.~\ref{fig:cold-shock}. Hence, in all the simulations, we perform the RT study after including the ionization effects from the shocks.

\begin{figure*}
 \centerline{
\def\arraystretch{1.0}
\setlength{\tabcolsep}{0.0pt}
\begin{tabular}{lcr}
    \includegraphics[width=0.33\linewidth]{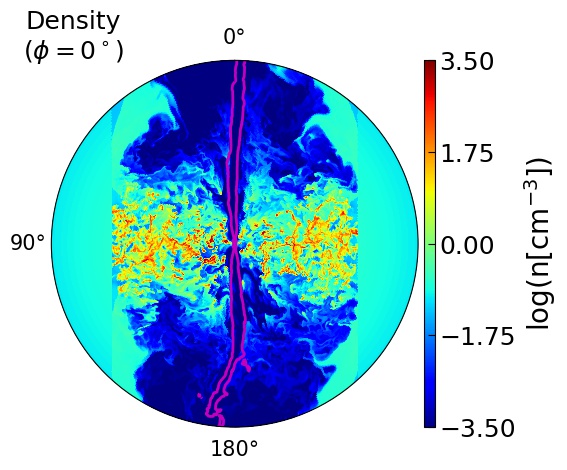} &
     \includegraphics[width=0.33\linewidth]{rho_t_90_70.jpeg} & 
    \includegraphics[width=0.33\linewidth]{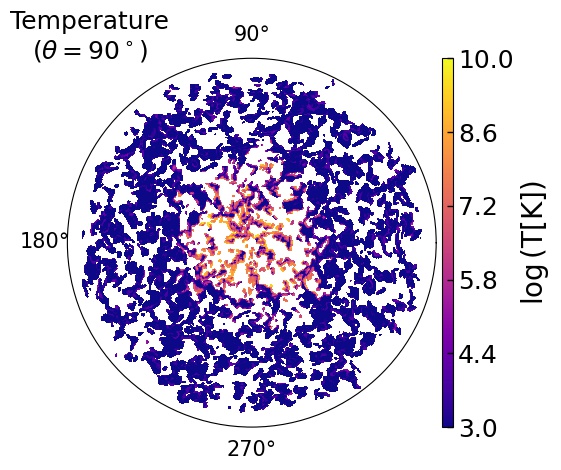}
     \end{tabular}}
\centerline{
\def\arraystretch{1.0}
\setlength{\tabcolsep}{0.0pt}
    \begin{tabular}{lcr}
     \hspace{-0.2cm}
     \includegraphics[width=0.3\linewidth]{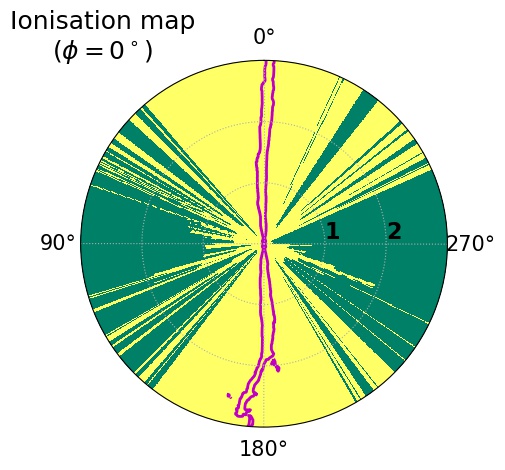} & \hspace{0.3cm}
    \includegraphics[width=0.31\linewidth]{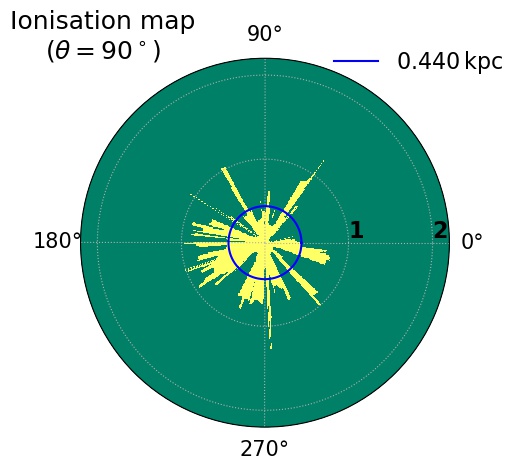} \hspace{0.1cm}
    \includegraphics[width=0.35\linewidth]{rho_t_90_70_shock.jpeg} &
     \end{tabular}}
     \caption{Density, temperature, and ionization maps for Sim.~B ($\theta_\mathrm{J}=0^\circ)$ at 0.68~Myr. \textbf{First Row:} Left: Density ($\log \mathrm{n[cm^{-3}]}$) slice in the vertical ($\phi=0^\circ$) plane (Extrapolating the cells using the initial halo profile (at $t=0$) results in the sharp variations in the density of the vertical plane), Middle: Density in disc ($\theta=90^\circ)$ plane, Right: Temperature ($\log \mathrm{T[K]}$) of the clouds in the disc plane. We use a cloud tracer of 0.99 to eliminate regions dominated by diffuse ionized gas from the temperature plot. \textbf{Second row:} Left and Middle: Ionization maps showing photo-ionized regions (in yellow) in the vertical and disc plane, respectively. The blue circle denotes the mean ionization radius in the disc plane (presented as $R$ in text). Right: Density distribution of unshocked gas in the disc plane. The magenta contour denotes the jet tracer at a value of 0.5 projected on the vertical plane. (The maximal value of the jet tracers is 1.)}
\label{fig:vert-70}
\end{figure*}

\subsection{Ionization due to the AGN and the jet}
\label{ion-jet-AGN}

\subsubsection{\textbf{`No-jet' simulation}}
\label{no-jet}
 For the `no-jet' (or control) simulation, we perform the RT study for the disc at 0.156~Myr. This snapshot corresponds to the unperturbed gas in the galaxy before the jet is ejected. 
The corresponding logarithmic density in the vertical ($\phi=0^\circ$) and disc ($\theta=90^\circ$) planes are presented in the first row of Fig.~\ref{fig:no-jet}. The density plots show an inhomogeneous disc of thickness $\sim 1$~kpc, and extending up to $\sim 2$~kpc in radius. We show temperature maps of the dense clouds in the disc plane (third panel in the top row of Fig.~\ref{fig:no-jet}), where the dense regions are identified using a threshold cloud tracer value of 0.99\footnote{The tracers in \textsc{Pluto} are passive scalars which are advected with the fluid. We use two types of tracers in our study,  `cloud tracer' and `jet tracer'. The cloud tracer follows the initial dense gas and the jet tracer follows the jet plasma is injected at the center and propagating through the computational domain.}.
The corresponding ionization map in the disc plane shows that AGN radiation can ionize just the inner region of the galaxy, giving a mean ionization radius of $\sim 0.18$~kpc in the disc plane. In the vertical plane ionization bi-cones are formed which are limited to a small angular width.

To investigate the variation of the extent of photoionized regions as a function of the polar angle ($\theta$), we compute the fractional obscuration of the photoionized radiation for a given $\theta$. An LOP is considered obscured if the ionization radius does not reach the end of the simulation domain because the AGN radiation is absorbed by the intervening gas. The obscuration fraction at a given $\theta$ is obtained by taking the ratio of the number of obscured LOPs to the total number of azimuthal LOPs. The results for different simulations are presented in Fig.~\ref{fig:jet-disc}.
The `no-jet' simulation shows strong shielding of the AGN radiation with only a small angular region near the poles being ionized (see Fig.~\ref{fig:jet-disc}), which forms the classic ionization bi-cones in Fig.~\ref{fig:no-jet}. This indicates that the AGN radiation can escape above and below the disc through diffuse regions close to the poles and absorbed elsewhere due to large column depths.

We tabulate the fraction of ionized region in the galaxy due to shocks and AGN SED in Table~\ref{tab:tab3}, which shows that for the `no-jet' simulation the fractions of ionized and dense ionized gas mass are 0.23$\%$ and 0.14$\%$ respectively. This indicates that in the absence of a jet, the AGN radiation cannot advance to large distances in the disc and has a negligible impact on ionizing the extended gas in the galaxy. In the next section, we discuss the ionization effects in the jetted simulations, where the jet is introduced at 0.17~Myr in the galaxy.

\begin{table*}
	\begin{center}
     \caption{Extent of collisional-ionization and photo-ionization in the gas discs.}
	\label{tab:tab3}
		\begin{threeparttable}
	\begin{tabular}{|c|l|l|l|l|l|l|l|l|l|l|} 
		\hline
     Simulation & &\multicolumn{1}{|c|}{No-jet} & & \multicolumn{1}{|c|}{B ($\theta_\mathrm{J}=0^{\circ}$)} & & \multicolumn{2}{|c|}{D ($\theta_\mathrm{J}=45^{\circ}$)} & &\multicolumn{1}{|c|}{E ($\theta_\mathrm{J}=70^{\circ}$)} \\  \hline
		Time of snapshot &  & 0.156~Myr && 0.68~Myr && 0.98Myr & 2.31~Myr &&  1.24~Myr \\
		 $R$ & & 0.18~kpc && 0.44~kpc && 0.736~kpc & 0.955~kpc && 0.933~kpc \\
		$M^{*}/M^\mathrm{T}$ & &  0.566 &&  0.583  && 0.599 & 0.692 && 0.6494 \\
		$M\mathrm{\SPSB{T}{sh}}/M^\mathrm{{T}}$  & &  0.154 && 0.279   && 0.3508 & 0.314 && 0.44   \\ 
		$M\mathrm{\SPSB{*}{sh}}/M^{*}$  & &  0.0034 && 0.161   && 0.25 & 0.181 && 0.331   \\ 
	    $M\mathrm{\SPSB{T}{ph}}/M^\mathrm{T}$ & &  0.0023 && 0.0124   && 0.024  & 0.033  && 0.038 \\ 
        $M\mathrm{\SPSB{*}{ph}}/M^{*}$  & & 0.0014 &&  0.019  && 0.0334  & 0.0416   && 0.052 \\ 
        FWHM (obscuration=0.5) & &  $158.4^\circ$ &&  $86.7^\circ$ &&  $68.8^\circ$ &  $62.6^\circ$ &&  $61.8^\circ$ \\\hline
	\end{tabular}
	\begin{tablenotes}
     \item $R$: Mean ionization radius in the disc plane. The distribution of the ionization radii for different simulations is discussed in Sec.~\ref{dist_radi}.
    \item $M^{*}/M^\mathrm{T}$: The dense gas phase mass fraction (defined as: $n>100\,\mathrm{cm^{-3}}$) in the galaxy, which traces the potential star forming regions,  \citep{dipanjan2018,mandal_2021}
    \item $M\mathrm{\SPSB{T}{sh}}/M^\mathrm{T}$: The fraction of total mass collisionally ionized by the jet-induced shocks.
    \item $M\mathrm{\SPSB{*}{sh}}/M^{*}$: The fraction of dense mass collisionally ionized by the jet-induced shocks.
    \item $M\mathrm{\SPSB{T}{ph}}/M^\mathrm{{T}}$: The fraction of the total mass ionized by AGN radiation.
    \item $M\mathrm{\SPSB{*}{ph}}/M^{*}$: The fraction of dense mass ionized by AGN radiation.
     \item FWHM : Full width at half maximum of the polar angle $\theta$, measured at an obscuration fraction of 0.5 for each simulation (see Sec.~\ref{ion-jet-AGN} for details).
  \item[] For `no-jet' case, the shocked mass includes the high temperature gas regions in the settling disc.
    \end{tablenotes}
    	\end{threeparttable}
    \end{center}
\end{table*}

\subsubsection{\textbf{Vertical jet}}
\textbf{Sim.~B ($\theta_\mathrm{J}=0^\circ$)}:
The jet in Sim.~B is ejected along the minor axis, and it breaks out easily at an early stage ($\sim$ 0.4~Myr) without significantly interacting with the disc. The escaping jet creates a high-pressure ellipsoidal bubble driving a strong shock through the ambient medium \citep{dipanjan2018}. We perform the RT study at 0.68~Myr when the jet plasma has reached the end of the simulation box. The density, temperature, and ionization maps for this snapshot are shown in Fig.~\ref{fig:vert-70}.
We use a value of 0.5 for the projected jet tracer in maps of vertical planes, which is shown with magenta contour. For the cloud tracer, we take a lower threshold value of 0.99 to identify the dense clouds, and show their temperature in the disc plane. One can see that after excluding the shocked gas (as described in Sec.~\ref{shocks}), a circular cavity of ionized gas is formed in the center of the disc plane. The RT study of the unshocked gas shows that the AGN radiation ionizes only the central region in the disc plane, giving a mean ionization radius of 0.44~kpc, while dense regions at a greater distance are unaffected.

In the ionization maps of the vertical plane, one can clearly see two ionization cones above and below the disc closer to the polar axis. While escaping from the disc, the vertical jet clears the dense gas  along its path, and hence the ionization cones are wider than those of the `no-jet' simulation (see Fig.~\ref{fig:no-jet}). Fig.~\ref{fig:jet-disc} shows that the region closer to the disc plane is obscured for a much smaller angular extent in Sim.~B (FWHM $\sim87^\circ)$ than the `no-jet' case (FWHM $\sim158^\circ)$. The obscuration fraction falls below $\sim 20\%$ for angles lower than $\lesssim 30^\circ$ from either poles. This implies a larger opening angle of the ionization cones, as the jet clears or shock-heats some gas from its immediate vicinity, opening up the path for the AGN radiation to penetrate to lower latitudes.

\begin{figure}
\centering
\includegraphics[width=0.9\linewidth]{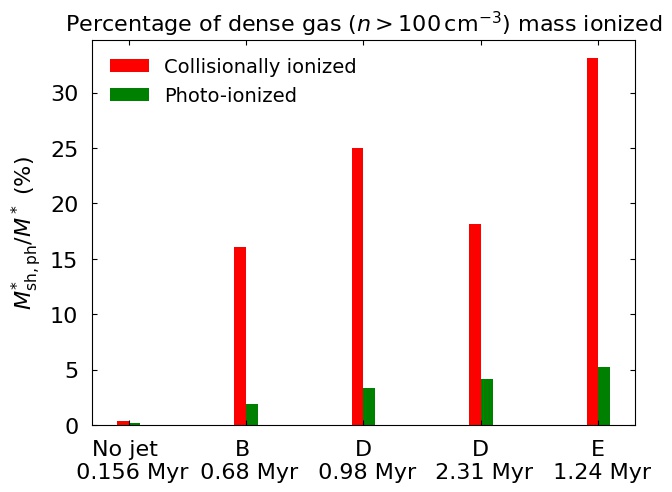}

\caption{The mass fractions of collisionally ionized and photo-ionized dense gas in different simulations. The fraction of ionized gas masses is tabulated in Table~\ref{tab:tab3}.}
\label{fig:dense_mass}
\end{figure}

From Table~\ref{tab:tab3}, the shocks from the vertical jet are found to ionize almost 16$\%$ of the dense gas mass ($\mathrm{n\,>100\,cm^{-3}}$) in the galaxy. The AGN radiation, on the other hand, ionizes only 1.9$\%$ of the gas mass in the galaxy, a significantly smaller fraction than the shocks. However, the ionization maps in Fig.~\ref{fig:vert-70} show that due to regions being cleared out by the jet, the radiation is able to propagate to large distances in the ISM (see Fig.~\ref{fig:dense_mass}).
This implies that photoionizing radiation from the AGN primarily ionizes  regions of diffuse gas in the galaxy, leaving the dense clumps unaffected. 

\textbf{Effect of Soft-Excess:\,}To estimate the contribution of the soft-excess in the ionization, we perform an RT study in the disc plane for Sim.~B after removing this component from the SED. This gives a mean ionization radius of 0.436~kpc in the disc plane at 0.68~Myr, which is very close to the mean value of 0.44~kpc when RT is performed using the full SED (see Fig.~\ref{fig:cold-shock}). The soft-excess \citep[modeled by using a black-body with $k_BT=85$][]{laha} contributes a luminosity of 1.5$\times 10^{43}\,\mathrm{erg\,s^{-1}}$ in the incident continuum  (see Fig.~\ref{fig:sed}), which is about 2 orders of magnitude less than the total incident luminosity, and hence is not expected to significantly affect the ionization in the galaxy.
This indicates that the contribution from the soft excess in the AGN SED used in our study is negligible in ionizing the disc gas.

\textbf{Ionization effects for more luminous RLAGN:\,} To check the impact of an AGN of higher power, we have performed an RT study for an incident spectrum of luminosity $10^{46} \ergs$.  The mean ionization radius for this case was found to be 0.475~kpc in the disc, a $9\%$ increase from that of the previous case. Thus the ionizing radiation penetrates  further into the ISM only by a small fraction, for an order of magnitude increase in the AGN power for the given gas distribution. Hence, we conclude that for the given gas distribution, the AGN radiation is unable to penetrate into the dense gas and the outer region of the disc remains unaffected from the AGN radiation of a highly luminous RLAGN as well, which is also shown in \citet{roos}.

\textbf{Effect of dust grains in the ISM on the photo-ionization:}\\
To examine how the grains in the ISM affect the extent of photo-ionization, we perform an RT run after excluding grains in the disc plane for Sim.~B. This increases the ionization radius to 0.48~kpc, which is only 10$\%$ greater than the previous value of 0.44~kpc. The grains in \textsc{Cloudy} are heated via absorption of incident radiation, by line and continuum emission within the clouds and gas collisions, and mainly cools by collisions with the gas and the thermionic effects. The temperature of the grains is determined by balancing the heating and cooling. We find that the grain temperature remains below 700~K for both graphite and silicates, which is lower than their sublimation temperatures of 1750~K and 1400~K for graphite and silicate, respectively. Also, the photoelectric heating due to the grains contributes up to $30\%$ of the total heating in most of the regions with dense ionized gas. This indicates that grains in the ISM negligibly affect the extent of photo-ionization.

 \begin{figure*}
 \centerline{
\def\arraystretch{1.0}
\setlength{\tabcolsep}{0.0pt}
\begin{tabular}{lcr}
      \includegraphics[width=0.33\linewidth]{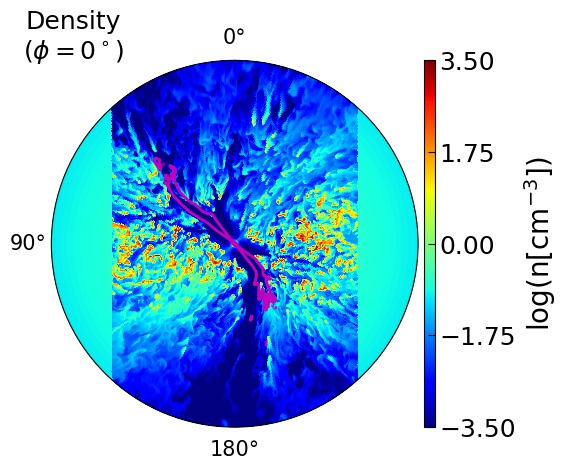} &
     \includegraphics[width=0.33\linewidth]{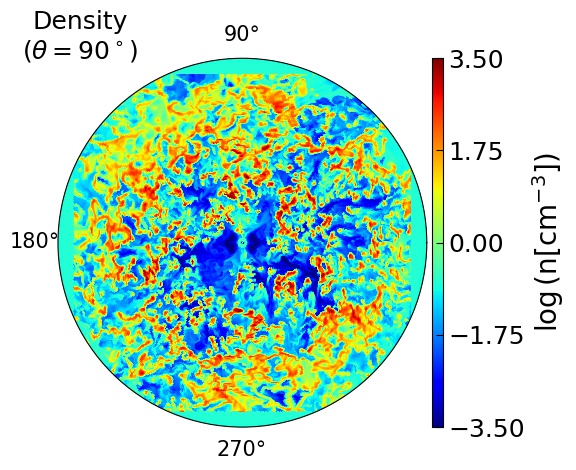} & 
    \includegraphics[width=0.33\linewidth]{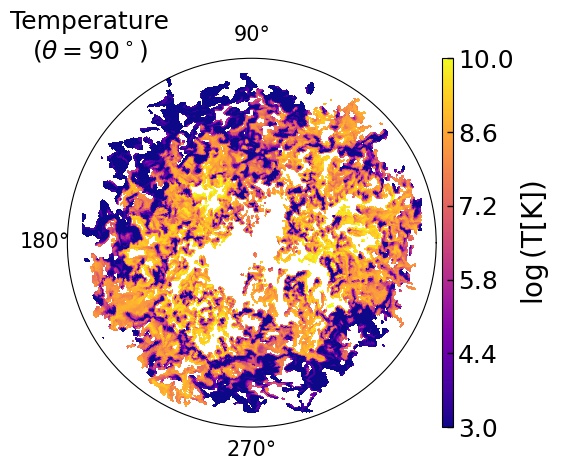}
     \end{tabular}}
\centerline{
\def\arraystretch{1.0}
\setlength{\tabcolsep}{0.0pt}
    \begin{tabular}{lcr}
     \hspace{-0.2cm}
     \includegraphics[width=0.3\linewidth]{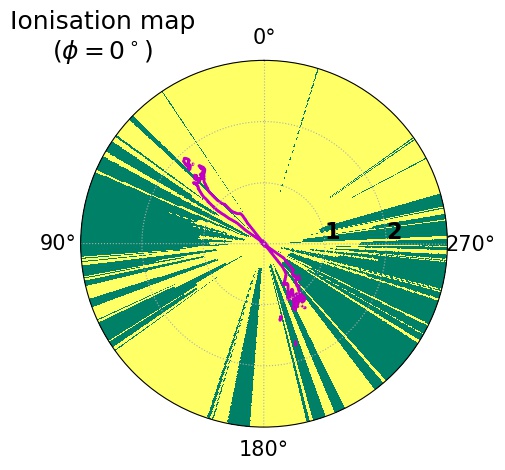} & \hspace{0.3cm}
    \includegraphics[width=0.31\linewidth]{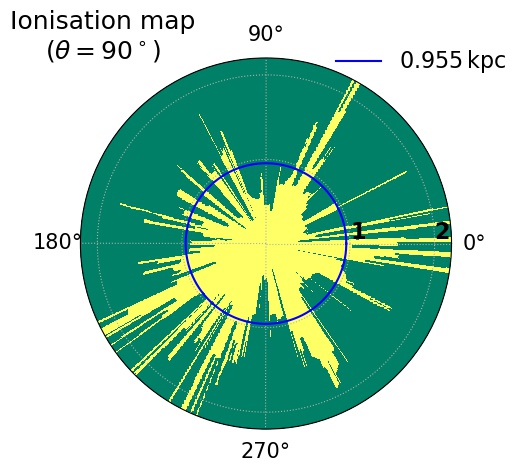} & \hspace{0.1cm}
    \includegraphics[width=0.35\linewidth]{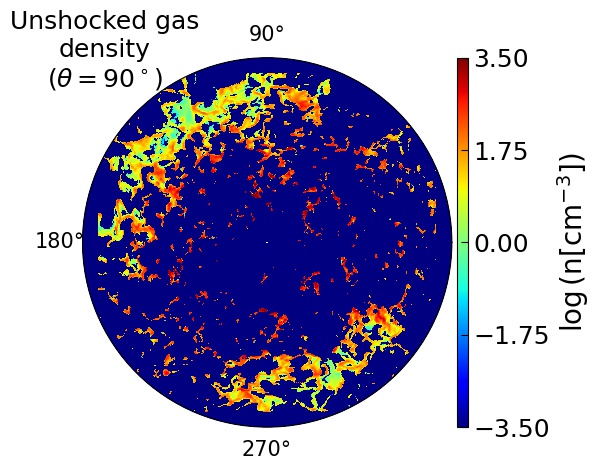} 
     \end{tabular}}
     \caption{Same as Fig. ~\ref{fig:vert-70} for Sim.~D ($\theta_\mathrm{J}=45^{\circ}$) at 2.31~Myr. We use an ionization radius value of 2~kpc for the fully ionized LOPs in the disc plane. }
\label{fig:inc-236}

 \centerline{
\def\arraystretch{1.0}
\setlength{\tabcolsep}{0.0pt}
\begin{tabular}{lcr}
    \includegraphics[width=0.33\linewidth]{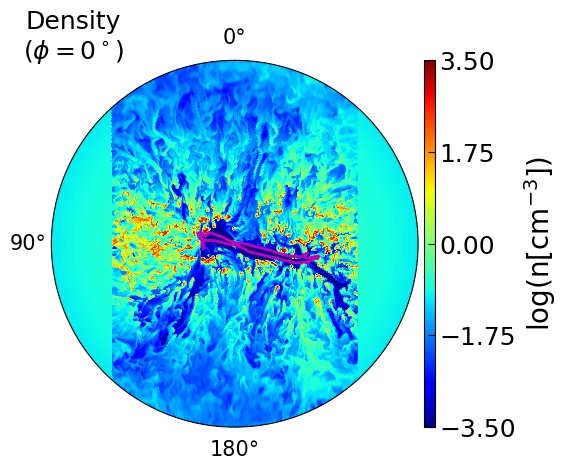} &
     \includegraphics[width=0.33\linewidth]{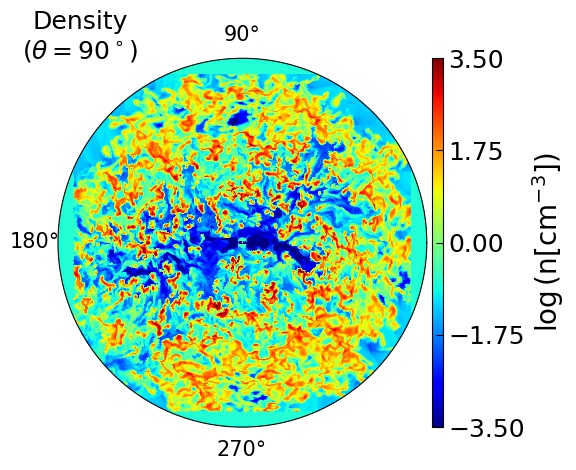} & 
    \includegraphics[width=0.33\linewidth]{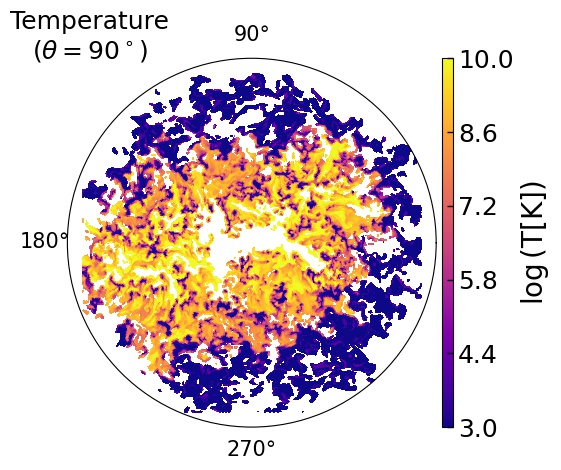}
     \end{tabular}}
\centerline{
\def\arraystretch{1.0}
\setlength{\tabcolsep}{0.0pt}
    \begin{tabular}{lcr}
     \hspace{-0.2cm}
     \includegraphics[width=0.3\linewidth]{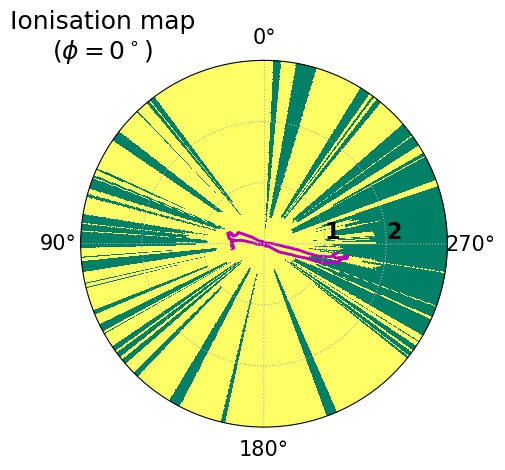} & \hspace{0.3cm}
    \includegraphics[width=0.31\linewidth]{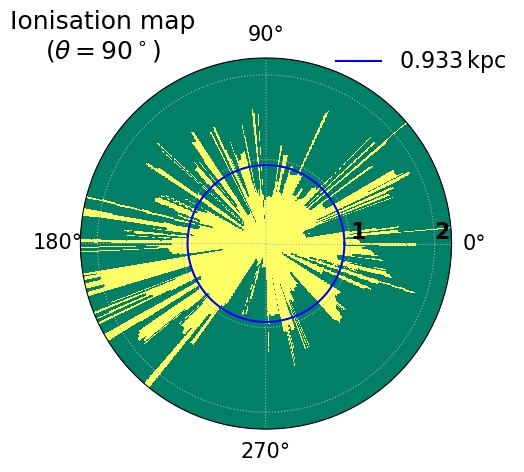} & \hspace{0.1cm}
     \includegraphics[width=0.35\linewidth]{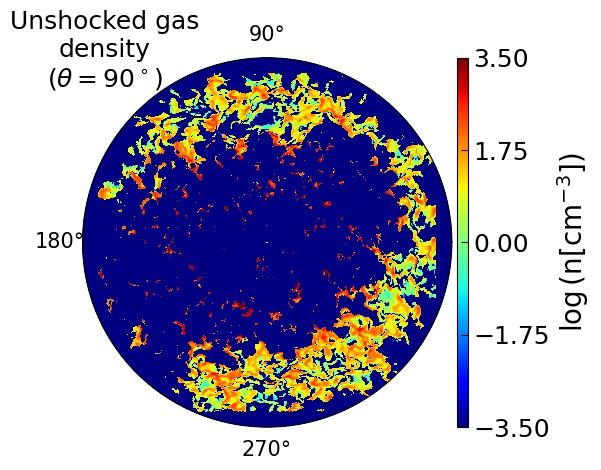} 
     \end{tabular}}
     \caption{Same as Fig. ~\ref{fig:vert-70} for Sim.~E ($\theta_\mathrm{J}=70^{\circ}$) at 1.24~Myr. We use an ionization radius value of 2~kpc for the fully ionized LOPs in the disc plane.}
\label{fig:inc-127}
\end{figure*}

\subsubsection{\textbf{Inclined jets}}
\label{inclined-jets}

In this section, we discuss the results of the ionization study in gas discs with jets inclined at an angle of $45^\circ$ (Sim.~D) and $70^\circ$ (Sim.~E). As compared to the vertical jets, the inclined jets are known to interact more strongly with the host's gas \citep{dipanjan2018}. Hence, in the sections below, we also compare the extent of jet-ISM coupling for vertical and inclined jets using various measures, including obscuration fraction, mean ionization radius and fraction of gas mass ionized in the disc. 
\begin{enumerate}
    \item 
\textbf{Sim.~D ($\theta_\mathrm{J}=45^\circ$):}
In Sim.~D, the jet is ejected at an angle of $45^\circ$ to the vertical axis. The different stages of evolution of the jet are shown in \citet[Fig.~14]{dipanjan2018}, where a spherical energy bubble is launched by the jet and drives forward shocks into the ambient medium. The bubble expands as the jet evolves into the ISM . 
 
\begin{itemize}
    \item \textbf{Morphology of the ISM:}
We perform the RT study for this simulation at two different times: i) at 0.98~Myr when the jet is still confined in the disc; and ii) at 2.31~Myr when the jet has broken free and evolved to larger scales. We show the density, temperature, and the ionized region of the simulation at 2.31~Myr in Fig.~\ref{fig:inc-236}. The unshocked gas is mainly distributed in the outer regions of the disc (bottom panel in Fig.~\ref{fig:inc-236}) as the jet has shocked the rest of the ISM in the central regions while propagating through it. The ionization map in the disc plane in Fig.~\ref{fig:inc-236} shows that the AGN radiation escapes easily from the low-density regions, ionizing some of the Lines of Propagation (LOP) up to the boudnary. The mean ionization radius in the disc plane is found to be 0.736~kpc at 0.98~Myr, and  0.955~kpc at 2.31~Myr respectively (see Table~\ref{tab:tab3}). This is significantly larger than the value of 0.44~kpc in Sim.~B (Fig.~\ref{fig:vert-70}), as the jet sweeps away more gas when inclined to the disc, due to stronger coupling with the ISM.\footnote{The comparison is done at 0.68~Myr for Sim.~B and 2.31~Myr for Sim.~E because the physical extent of the jet is similar at these snapshots \citep{dipanjan2018}.}

\item \textbf{Wider ionization cones:}
We find wider ionization cones in the vertical plane, than those found for the vertical jet (Sim.~B). 
As shown in the vertical plane ionization map of Fig.~\ref{fig:inc-236}, some dense clouds near the central regions cast shadows, identified as streaks of unionized LOPs between fully ionized regions. 
Fig.~\ref{fig:jet-disc} shows that more penetration of the AGN radiation into the disk results in lowering of the obscuration fraction for polar angles closer to the equator for Sim.~D. Also, the FWHM (width of polar angle at an obscuration fraction of 0.5) is decreased to $\sim63^\circ$ at 2.31~Myr than its previous value of $\sim69^\circ$ at 0.98~Myr (see Table~\ref{tab:tab3}). However, the late time snapshot of Sim.~D ($t=2.31$~Myr) shows a distinctly different feature in regions close to the southern pole ($\theta \sim 150^\circ - 180^\circ$) where the obscuration is seen to increase near the vertical axis. Such enhanced obscuration is not seen at the earlier time of the same simulation (blue dotted line, at $t=0.98$~Myr). This occurs as the escaping jet imparts momentum to the gas in the disc and lifts some of the clouds to higher altitudes above and below the disc. 
This may be attributed to the disc gas being shredded and pushed along the polar regions by the escaping jet. 
Such a disrupted ISM will result in partial obscuration of photoionization cones in Fig.~\ref{fig:inc-236}, as the uplifted clouds add to the optical depth along an LOP, resulting in dark patches inside the ionization cones. 

\item \textbf{Photo-ionization vs collisional ionization:}
We find that the mass fraction of dense gas increases with time for Sim.~D, by nearly $\sim 10\%$ from the value at 0.98~Myr (as shown in Table~\ref{tab:tab3}). This is due to the strong compression arising from radiative shocks driven by the inclined jet \citep{Sutherland07,dipanjan2018}.  The shocks from the jet ionize 18$\%$ of the potential star-forming mass in Sim.~D, comparable to that in the vertical jet (16$\%$). However, the fraction of collisionally ionized gas decreases between the two snap shots of Sim.~D, as shown in Table~\ref{tab:tab3}. As the jet decouples from the disc  at later times, the shocked gas in the disc thus cools down, leading to a decrease in the shocked gas fraction. 

The mass fraction of the total gas ionized by the AGN radiation mildly increase with time, as shown in Table~\ref{tab:tab3}. However, the dense gas remains largely unaffected, with a maximum of 4$\%$ of the dense mass fraction of Sim.~D being ionized by the AGN. This is again comparable to that of the vertical jet in Sim.~B ($1.9\%$). Similar to Sim.~B, although the shocks from the jet are powerful in clearing out the path for AGN radiation and ionizes a larger fraction of the dense mass in the disc, the AGN radiation affects the dense gas clouds only weakly. 

\end{itemize}

\item
\textbf{Sim.~E ($\theta_\mathrm{J}=70^{\circ}$):}
The jet in Sim.~E is ejected at an angle of $70^\circ$ to the vertical axis. The jet being almost parallel to the disc plane, encounters a high column depth of clouds and launches sub-relativistic outflows through channels into the disc \citep{dipanjan2018}.
We perform the RT study for Sim.~E at 1.24~Myr, and the plots for the density, temperature, and ionization in the vertical plane and the disc plane are shown in Fig. \ref{fig:inc-127}. The two lobes of the jets encounter different local gas distributions in the disc and thus interact differently with the gas, resulting in a crescent-shaped cavity in the disc plane. The corresponding ionization map shows that the AGN radiation can now escape through several shocked regions, resulting in a mean ionization radius of 0.933~kpc.

\begin{itemize}
\item \textbf{ Wide partially obscured ionization cones:}
In the ionization map of the vertical plane, one does not see the classical bi-conical structures of the ionization cones above and below the disk here, as several regions are unevenly shielded from the AGN radiation. As the jet progresses into the disc, it opens up several channels and re-distributes the dense clumps in the vertical directions. This creates several alternating regions of ionized LOPs, and ones that are obscured due to dense clouds. This morphology is different from Sim.~B and Sim.~D, where the ionization cones were broad and intact. The jets in such cases are more successful in clearing the gas along their path, and the AGN radiation can freely escape along the radial directions. 
In Sim. E, however, the jet is almost parallel to the disc, remains confined in the ISM, and directly interacts with the gas disc. This lowers the obscuration fraction for Sim.~E near the equatorial regions, and the disc is obscured for a smaller angular extent (see red curve in Fig.~\ref{fig:jet-disc}, FWHM $\sim62^\circ$), as compared to the vertical jet (FWHM $\sim87^\circ$).

\item \textbf{Ionization of dense gas:}
The ionization study for the disc in Table~\ref{tab:tab3} shows that the shocks from the jet ionize up to $33\%$ of regions of dense mass. Since the jet is nearly inclined into the dense gas in the disc, it is capable of affecting a larger portion of the ISM. This allows the AGN radiation to propagate to large-distances in the disc, as visible from the ionization maps in Fig.~\ref{fig:inc-127}; however, only $5.2\%$ of the dense mass in the galaxy is photo-ionized.
Thus similar to previous cases, the AGN radiation seems ineffective in ionizing the dense gas in the galaxy, as shown in Fig.~\ref{fig:dense_mass}). It is mostly the diffuse gas regions that are being affected, and the dense clumps remain shielded from the AGN radiation. 
\end{itemize}

\subsubsection{\textbf{Ionization radii in different simulations}}
\label{dist_radi}

\begin{figure}
\centering
\includegraphics[width=0.7\linewidth]{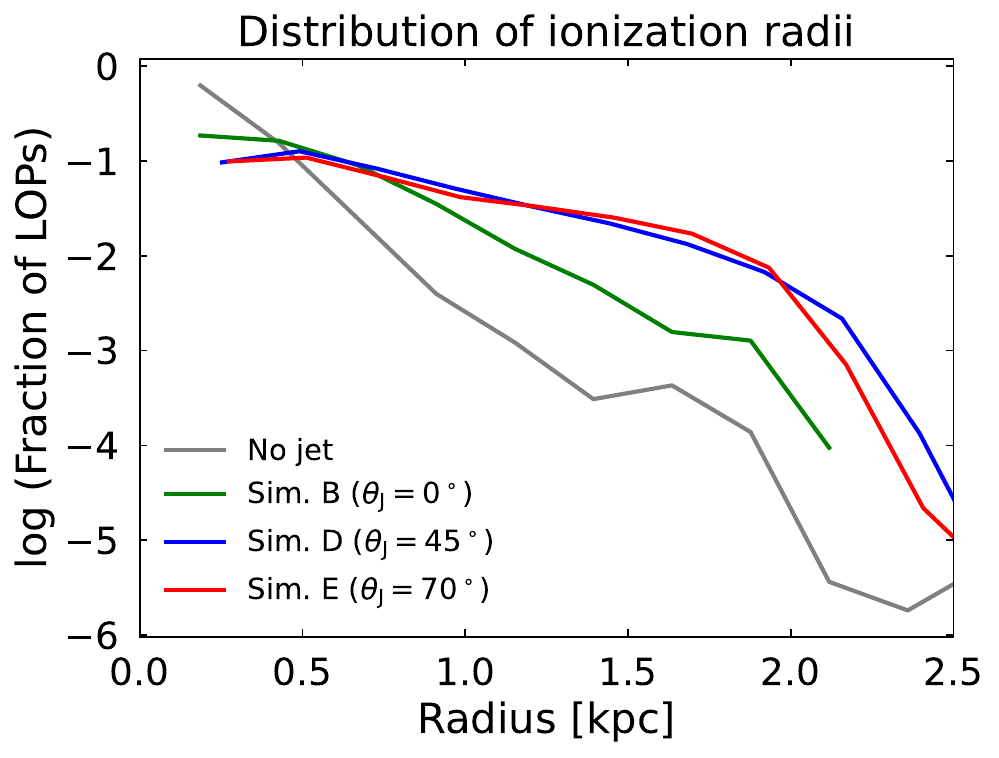}

\caption{Distribution of ionization radii in different simulations.}
\label{fig:1d_hist}
\end{figure}

\begin{figure*}
\hspace{-1cm}
 \centerline{
\def\arraystretch{1.0}
\setlength{\tabcolsep}{0.0pt}
\begin{tabular}{lcr}
      \includegraphics[width=0.35\linewidth]{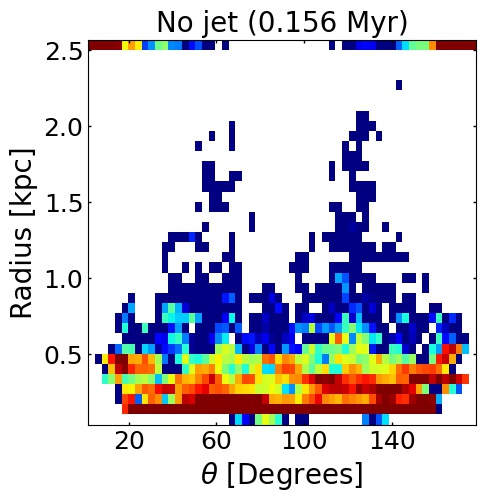} &
     \includegraphics[width=0.33\linewidth]{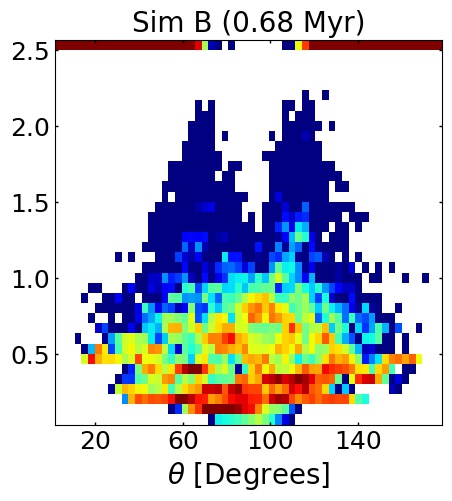} & 
    \includegraphics[width=0.389\linewidth]{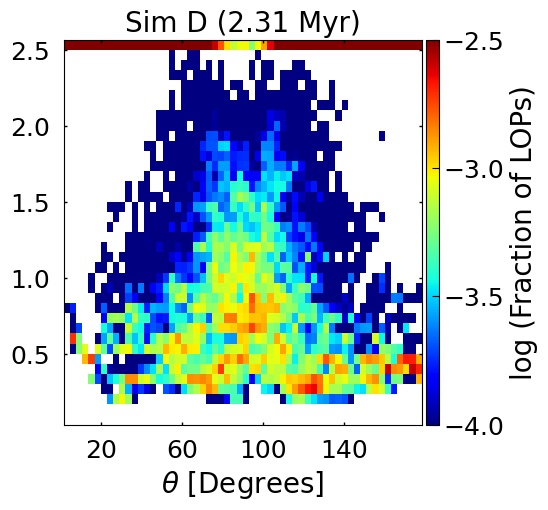}
     \end{tabular}}
     \caption{\textbf{Left to right: } Distribution of ionization radii as a function of $\theta$ for `no-jet', Sim.~B ($\theta_\mathrm{J}=0^\circ$) and Sim.~D ($\theta_\mathrm{J}=45^\circ$). In the above plots, the ionization radius for fully ionized LOPs (i.e. $R\approx5~$kpc) is replaced with 2.5~kpc, which appear as the large bins accumulated at the top.}
\label{fig:2d_hist}
\end{figure*}

    In this section, we discuss the distribution of the ionization radii in different simulations. We plot the 1-D distribution of the ionization radii and its variation as a function of the polar angle $\theta$ for different systems in Fig.~\ref{fig:1d_hist} and Fig.~\ref{fig:2d_hist}, respectively. These plots show that the ionization radii for obscured LOPs have a wide range of values, from 0.5~kpc to 2~kpc.
    In the `no-jet' case, the ionization radii for a significant fraction of the LOPs is less than 0.5~kpc, and only a few are ionized to large distances in the disc i.e., $R$~>~2~kpc. However, in the jetted simulations, several LOPs even at large angles ($\theta$) from the poles are fully ionized. Also, the fraction of LOPs ionized to large distances ($\sim2$~kpc) in the jetted simulations is higher when compared with the `no-jet' case.
    Sim. B, with the jet vertical to the disc, has a large fraction of ionization radii within 0.5 kpc of the disc plane ($20^\circ \lesssim \theta \lesssim 140^\circ$). However, some LOPs also show higher ionization radii ($\sim 2$) kpc at $\theta \sim 60^\circ$ and $\sim 120^\circ$, which mark the edges of the gas disc being cleared by the outflow. Radiation can penetrate to larger distances at such angles. On the other hand, Sim.~D, with the jet inclined to the disc by $\theta_\mathrm{J} = 45^\circ$, shows a much wider distribution of ionization radii in the disc mid-plane, as the jet clears the gas for the ionizing radiation to penetrate. The above also explains the lower obscuration fractions in these regions, as described previously in Fig.~\ref{fig:jet-disc}.

\subsubsection{\textbf{Evolution of the mean ionization radius with time for different simulations:}}
To investigate the evolution of the extent of the photoionized regions with time in the disc plane, we present in Fig.~\ref{fig:ion-rad} the time evolution of the mean ionization radius in the disc plane for all the simulations in Table~\ref{tab:sim_table}.  For the `no-jet' simulation, the ionization radius is small, and remains steady around $\sim 0.14$~kpc as the mean column depth of the gas in the disc varies only mildly as the disc settles. 
In the jetted simulations, the ionization radius increases with time since the jet sweeps away gas while interacting with  the disc.
At a given time, higher jet inclination leads to increased mean ionization radius. The inclined jets clear out more gas in the disc due to stronger coupling with the ISM (see density in disc plane in Fig.~\ref{fig:inc-236} and Fig.~\ref{fig:inc-127}). This leads to more of the central region being ionized by the AGN radiation. However, we note here that as discussed previously, the AGN's radiation is weak in significantly ionizing the dense cores of the nuclear disc regions (up to $\sim1$~kpc). Thus, the increase in ionization radius is not expected to affect the star formation rate in gas disc.

\end{enumerate}
\begin{figure}
  \centering
{\includegraphics[width=0.8\linewidth]{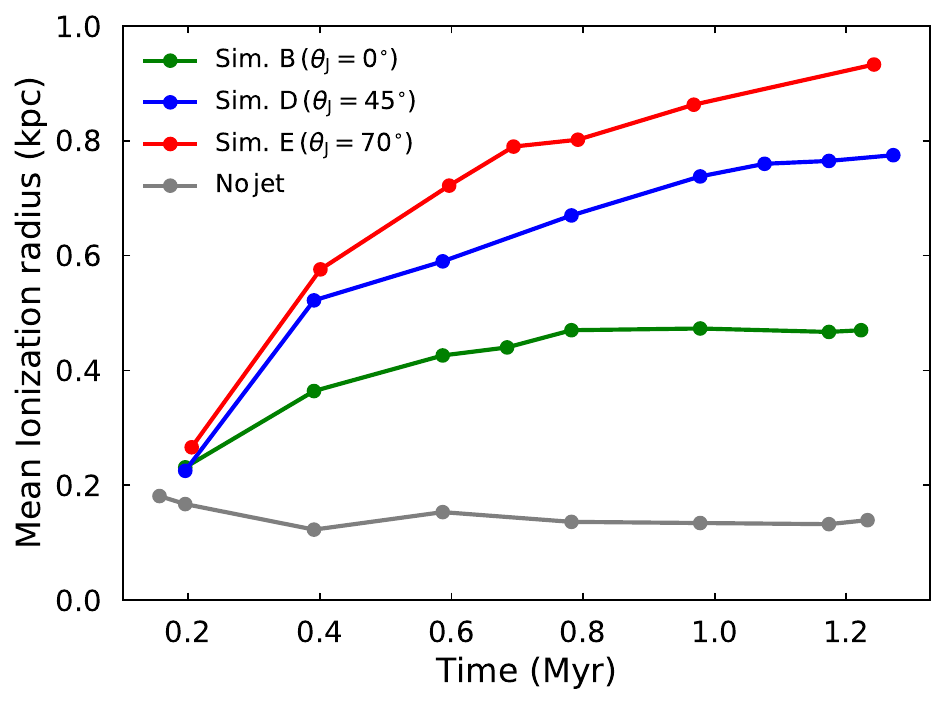}}
\caption{Time evolution of mean ionization radius in the disc plane ($\theta=90^\circ$) for simulations in Table~\ref{tab:sim_table}.}
  \label{fig:ion-rad}
\end{figure}

\section{Discussion}
\label{discussion}
In this study, we have addressed how the radio-loud AGNs can ionize the central few kpcs of their host's ISM, mediated either by the photo-ionization from the AGN's radiation or through shocks driven by the jet-ISM interaction. We also explore the extent of dense gas mass ($n>100\,\mathrm{cm^{-3}}$) ionized by the AGN in the disc, which constitutes the potential star-forming regions. However, to examine the effect of jets or radiation pressure on the star-formation activity in addition to ionization, a thorough treatment of the  turbulence-regulated star-formation efficiency  \citep[as done in][]{mandal_2021} is required, but here we use a simplified analysis and only look at how much of the potential star-forming dense gas is ionized. Our study is similar to the photo-ionization study by \citet{roos}, except that the galaxies in this work host powerful jets launched at different angles to the disc plane. The previous study was performed on a large-scale disc impacted by an AGN-driven wind, which was found to be weak in affecting the star-formation activity. In the following sections we discuss some implications that can be drawn from our study.

\begin{figure}
\centering
\includegraphics[width=0.7\linewidth]{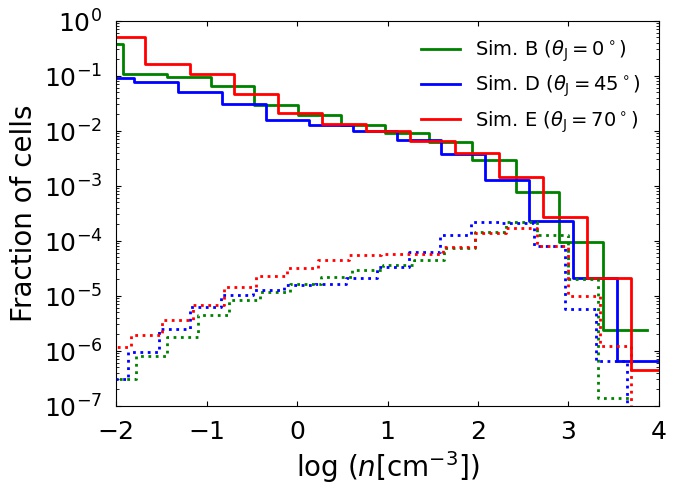}
\caption{Density distribution of collisionally ionized (solid) and photo-ionized (dotted) gas in different simulations.}
\label{fig:hist-jet-rad}
\end{figure}

\begin{enumerate}
\item \textbf{Ionization from AGN SED:}
In the Radiative Transfer (RT) study using \textsc{Cloudy}, we find that the AGN radiation is not able to affect a substantial fraction of the dense gas ($n>100\,\mathrm{cm^{-3}}$) in the disc.
Similar results have also been obtained by \citet{roos}, where it was found that the self-shielding of the dense regions makes it difficult for the ionizing radiation from the AGN  to penetrate the clouds. Although we find that the AGN radiation can heat up the outer layers of dense regions to higher temperatures than the cooling floor of the simulation, the mass filling fraction of such regions is very low (see Appendix~\ref{rad-jet-heat}). Thus, we don't expect AGN heating to be substantial in affecting the dynamics and star-forming activity in the galaxy. However, one should note that we probably underestimate the true extent of photo-ionized regions, because the RT computations in \textsc{Cloudy} do not include the effects of radiation pressure on the gas dynamics which can assist in dispersing dense gas, as discussed in Appendix~\ref{mom-compare}.

\item \textbf{Effect of jet inclination on the ionization of the ISM:}
Our analysis shows that the presence of jet-ISM Interaction can have a significant impact on the gas distribution of the disc. The jets can clear out the gas in the nuclear regions and make paths for AGN radiation to escape, as compared to the no-jet system, where the AGN radiation can only escape through the diffuse regions above and below the disc and is shielded elsewhere. 
We also find that, compared to the vertical jets, the inclined jets can ionize a larger fraction of dense gas in the disc (see Table~\ref{tab:tab3} and Sec.~\ref{inclined-jets}). The more inclined jets remain confined for longer times and exhibit stronger jet-disc interactions \citep{dipanjan2018}, which enables them to shock-heat more gas along their path.
These jets heat up and push out a substantial amount of gas mass in the central disc regions as they progress, enabling the AGN radiation to ionize gas further out in their hosts, compared to the vertical jets which escape easily without much interaction (see Fig.~\ref{fig:ion-rad}). This also lowers the fraction of Lines of Propagation (LOPs) that are obscured by dense gas closer to the equatorial regions of their systems (see Fig.~\ref{fig:jet-disc}).
However, the inclined jets can also uplift the gas in the vertical direction, which can  increase the obscuration fraction in the regions closer to the poles due to shielding of AGN radiation by the entrained clouds.

\item \textbf{Radiation vs jets on ionization of gas:}
As discussed earlier, we find that the jets are able to ionize a significant fraction of dense-gas mass in the discs, whereas the cumulative effect of AGN radiation is comparatively very weak (see Fig.~\ref{fig:dense_mass}). Below we compare the extent of ionization for both these mechanisms on the density distribution in the discs. In Fig.~\ref{fig:hist-jet-rad}, we show the fraction of cells that are collisionally and the fraction that are photo-ionized as a function of density in different simulations. The figure demonstrates the following points:
\begin{itemize}
    \item We firstly note that the distribution of collisionally ionized gas fraction peaks at lower densities, with a fall off at higher values. This occurs as shocks significantly slow down at high densities and hence cannot efficiently reach the inner cores (see \citet[Fig.~3]{mandal_2021} for PDF of density in the Sim.~D). However, the density distribution of photo-ionized cells in Fig.~\ref{fig:hist-jet-rad} peaks at high values ($n\sim 100\cc$). This demonstrates that although shocks may not efficiently penetrate the dense cores, the AGN radiation can, as it depends on the total column depth along a given ray.
    \item
    We however still note that the total fraction of photo-ionized cells is lower than the collisionally ionized case. This shows, that AGN radiation  is not able to affect a significant overall fraction of total dense gas, although it is more penetrative; however, this can also result due to radiation being facilitated by the jets by ionizing the regions in front of the dense cells.
    \item 
    Both radiation and shock heating are weak in the densest ($\sim10^{3}~\mathrm{cm^{-3}}$) regions of the host's ISM.
    \item One can notice that the distribution of densities affected by the jets in Fig.~\ref{fig:hist-jet-rad} are almost similar for different jetted simulations. This implies that, whether clouds are hit by the main jet stream, secondary jet streams, or backflows, may not affect the way clouds are shocked, as long as they are being hit sufficiently strongly by the jet flows.
\end{itemize}

\item \textbf{Ionization cones in the jetted systems:}
We find that the escaping jets (Sim.~B ($\mathrm{\theta_J=0^\circ}$) and Sim.~D ($\mathrm{\theta_J=45^\circ}$)) clear out the gas in the vertical direction, which enables the AGN radiation to propagate freely in the regions above and below the disc. This produces much  wider ionization cones in these simulations (see Fig.~\ref{fig:vert-70} and Fig.~\ref{fig:inc-236}), as compared to the `no-jet' disc, although one can see some dark streaks in the lower ionization cone for Sim.~D (close to $\theta=180^\circ$) due to increased obscuration fraction in the these regions (see Fig.~\ref{fig:jet-disc}).
Such wide bi-conical ionization structures have been observed in several Seyfert galaxies, such as NGC~4388 \citep{corbin_1988}, NGC~1068 \citep{pogge}, NGC~5252 \citep{tadhunter_1989} and NGC~5728 \citep{wilson}.

However, for Sim.~E ($\mathrm{\theta_J}=70^\circ$), the jet, while propagating into the disc, creates several channels and uplifts the gas in the regions above and below the disk (see Fig.~\ref{fig:inc-127}). This creates non-classical bi-conical structures in the vertical plane that appear as dark filaments when seen against the photoionized cones. These filaments are produced by the shadows of the dense clumps, which shield these regions from the ionizing radiation of the AGN.
Such contrasting features of bright and dark rays in [\ion{O}{III}] emission have been observed in the nearly edge-on viewed galaxy IC~5063 \citep{peter20}. A similar mechanism, where irregular obscuration of the central AGN due to a jet disturbed ISM could be at play here. 
However, we must note that Fig.~\ref{fig:inc-127} shows the ionization region for only one plane (viz. the vertical plane, $\phi=0^\circ$), and an observer will see an integrated image containing the superposition of such planes as the observer's LOS passes through the galaxy. This may dilute the sharp contrasting features in the photo-ionized and shielded regions. The shadows may further be diluted by diffuse emission from the surrounding gas, which was not considered here. 

\item \textbf{Winds vs jets on ionization in the host:}
In our study, we find that the jet-induced shocks can collisionally ionize a substantial fraction of the gas in the central regions ($\sim 2 \kpc$) of the host galaxy, as shown in Table~\ref{tab:tab3}. This is similar to the findings of \citet{gabor13} and \citet{roos}, which found strong heating from the AGN-driven wind in the nuclear regions of the large disc. However, these studies showed that the wind is unable to affect the gas up to large scales in the host. This may be due to the fact that AGN-driven winds are strongly hindered by the gas. In contrast jets can induce a large momentum in the host out to large radii, as they remain
collimated for longer times \citep{dipanjan_2021}.
One can see that the AGN luminosity in the simulations from \citet{gabor13} varies by six orders of magnitude $(10^{40}-10^{45}\ergs)$ on a time-scale of 100~Myr, as shown by \citet{gabor13}, but the change in star formation rate (SFR) is not substantial.
In contrast, the jets in the simulations used in our study are able to drive powerful shocks into the ISM during a short period of a few Myrs, affecting a large fraction of potentially star-forming dense gas.
However, we must note that the study by \citet{gabor13} was conducted on a bigger disc galaxy, roughly an order of magnitude greater in mass and radius than the disc in the simulations utilised in our work, and hence the direct quantitative comparison between the two setups may not be right.
However, one can conclude that the jets as well as winds can have a strong impact in the central regions of their host galaxy.

\item \textbf{Does AGN radiation mask the ionization from shocks?}\\
Photo-ionization from the AGN is found to dominate the ionization in the nuclear regions of several galaxies, whereas, the jets are found to affect the regions at large scales \citep{tadhunter_2002,holt_2009}.
In our study, we find that the jet-induced shocks can ionize the gas-disc to large scales (up to $2$~kpc) and clear out the gas in the nuclear region of the disc, which enables the AGN radiation to penetrate to large distances into the ISM (see Fig.~\ref{fig:cold-shock} and~\ref{fig:2d_hist}). However, the radiation seems to have a weak effect in ionizing the dense mass in the galaxy and is not likely to propagate to large distances without the contribution of shocks, as seen in the no-jet case (see Fig.~\ref{fig:no-jet}). One can expect that the shock-ionized gas due to the jet will begin cooling down after the jet stops interacting with a cloud, and the cooling gas behind the shocks is likely to be ionized by the AGN radiation. Since the cooling of the shocked gas is not an abrupt process, the gas density of such regions faced by the radiation may not increase the column density to high values to stop the radiation. Such post-shock gas that is ionized by the AGN is likely to produce line ratios characteristic of photo-ionized gas in different wavelength regimes, although its kinematics can still show signatures of jet interaction, as has been observed in the nuclear regions of several studies \citep{solorzano_2002,holt_2009,santoro_2020}.
However, this possibility needs to be investigated in detail by conducting a study with both jet and AGN radiation simultaneously active and by examining their contribution to the ionization of different regions in the galaxy.

\end{enumerate}

\section{Summary and Conclusions}
\label{summary}
In this study, we investigated the impact of ionization due to powerful jets and AGN radiation on kpc-scale gas-rich discs. A radiative transfer (RT) study was performed using \textsc{Cloudy} in post-process with data of the simulations from \citet{dipanjan2018}. For this, we constructed a radio-loud AGN SED consistent with the mean SED of \citet{shang}, as the central ionizing source. Before performing the RT study, we also identified the ``collisionally-ionized" gas cells and filled them with a very low gas density, as discussed in Sec~\ref{shocks}. The main findings of our study are summarised below.
\begin{itemize}
    \item In the `no-jet' (or control) simulation, the disc is clumpy, and most of the dense regions are at lower temperatures ($<10^4\, \mathrm{K}$). The RT study shows that most of the galaxy remains unaffected by the AGN's ionizing radiation, with less than $1\%$ of the dense gas mass being photoionized.
    \item In the jetted simulations, the jet drives strong shocks into the ISM, ionizing a significant fraction of dense gas (up to $33\%$). The shocks from the jet collisionally ionize the gas in the central region and thus, enable the AGN radiation to penetrate farther into the galaxy. However, the fraction of dense gas ($n > 100\, \mbox{cm}^{-3}$) mass affected by the AGN ionizing radiation is comparatively smaller (up to $5.2\%$), than what we obtained for the jet-induced shocks. 
    
    \item RT study in different jetted simulations indicates that AGN radiation can ionize only the diffuse gas regions in the galaxy, while a large fraction of dense gas patches is almost unaffected. This implies that the ionized radiation is unable to penetrate deep into to the dense gas clouds. Such self-shielding from the outer layers of the dense clumps blocks the AGN radiation and prevents it from affecting the inner regions, as has been shown earlier by \citet{roos} and \citet{bieri17}.
    
    \item As the jet evolves, it sweeps away more gas, and the mean ionization radius in the disc plane increases with the jet propagating into the ISM. Higher inclination of the jet leads to a greater ionization radius in the disc plane. However, since the dense regions are almost unaffected, the larger ionization radius does not necessarily imply a strong effect on the star-forming activity.
 
    \item We find that the angular variation of the obscuration fraction (fraction of obscured LOPs by dense gas) depends highly on the jet inclination angle and the stage of jet evolution in the disc. The vertical jet drills out easily from the disc without much interaction, and so the regions closer to the equator remains obscured. On the other hand, the inclined jets exhibiting jet-ISM coupling, affect the optical depth of the ionizing radiation at a given azimuthal angle differently. The jet at $45^\circ$ pushes the disc gas along the poles while escaping from the disc, leading to an increase in the obscuration fraction in regions closer to poles. In contrast, the jet ejected at $70^\circ$ remains confined into the disc, and vigorously interacts with the ISM, lowering the obscuration in the equatorial regions.

\end{itemize}

\section*{Acknowledgements}
We are grateful to the referee, Dr Salvatore Cielo, for his useful and constructive comments, which helped in improving the clarity of this manuscript. We also thank Gulab Dewangan and Raghunathan Srianand for helpful discussions regarding photo-ionization study in the disc using AGN SED. Moun Meenakshi also thanks Ankush Mandal for his assistance with the Python computations used in this study. The computations in this study were performed on the supercomputing facility at IUCAA\footnote{\url{http://hpc.iucaa.in}}, Pune. This research was undertaken with the assistance of resources from the National Computational Infrastructure (NCI Australia), an NCRIS enabled capability supported by the Australian Government. AYW is supported by JSPS KAKENHI Grant Number 19K03862. R.M.J. Janssen is supported by an appointment to the NASA Postdoctoral Program at the NASA Jet Propulsion Laboratory, administered by Universities Space Research Association under contract with NASA.


\section*{DATA AVAILABILITY}
No new data were generated in support of this work. The simulations used are available from the corresponding authors upon reasonable request.



\bibliographystyle{mnras}
\bibliography{manuscript} 

\appendix

\section{Momentum due to radiation and jet} 
To determine whether the radiation or jet governs the dynamics of the disc gas, we estimate the expected momentum injection rate for both in the ISM.
\label{mom-compare}\\
\textbf{Momentum due to radiation:} 
 The ionizing radiation ($L>1$ rydberg) in the AGN SED constitutes a luminosity of $10^{44.6}\mathrm{\ergs}$. This gives a momentum injection rate of $1.327\times 10^{34}\,\mathrm{dyn}$. Assuming that the outflow kinetic power for radiation driven winds is up to $5\%$ of the bolometric luminosity of the AGN \citep{zubavous_2012}, gives the momentum injection rate of $6.6\times 10^{32}\,\mathrm{dyn}$.

\textbf{Momentum due to jet:}
The kinetic momentum injection rate due to the jet is given as, $\Gamma \dot{M}_{\mathrm{jet}} \beta c$, where where $\Gamma$ is the jet Lorentz factor, $\dot{M}_{\mathrm{jet}}$ is the mass flux, and $\beta$ is the jet velocity in units of the speed of light. The mass flux (\citet{wagner12}) due to the jet is:
 \begin{equation}\label{eq:mom_jet}
 \dot{M}_{\mathrm{jet}}=\frac{1}{\Gamma -1}\frac{P_{\mathrm{j}}}{c^2}\left(1+\frac{\Gamma}{\Gamma-1}\frac{1}{\chi}\right)^{-1}\end{equation}

For the simulations used in this study (see Table~\ref{tab:sim_table}), $P_\mathrm{j}=10^{45}~\mathrm{erg\,s^{-1}}$ is the kinetic power of the jet, $\chi=10$, $\Gamma=(\sqrt{1-\beta^2})^{-1}=5$, and $\beta=0.979$. This gives the kinetic momentum injection rate from the jet to be $3.625\times 10^{34}\,\mathrm{dyn}$, which is approximately 55 times that from the AGN radiation estimated above. 
To further compare the affect of jet and radiation, we also estimate the momentum boost (or mechanical advantage) by jet which is the ratio of radially outward momentum and the total momentum injected by the jet. Using the momentum injection rate from the jet, we get a value for boost to be 493 at 0.68~Myr for Sim.~B, similar to the values obtained in \citet{wagner11,wagner12}. Contrarily, the AGN radiation-driven outflows can boost the momentum up to 20 times \citep{zubavous_2012,tombesi15}.\\
The momentum injected by AGN radiation for a highly luminous quasar of bolometric luminosity $\sim10^{47}\ergs$ will be comparable to that from the jet of power $10^{45} \ergs$ here. However, since the dense regions shields themselves from the ionizing radiation (\citet{bieri17}), and due to the computation being static in \textsc{Cloudy}, the affects of a luminous AGN SED in ionizing the ISM will not be significant, as shown in \citet{roos}.

\textbf{Momentum boost due to IR trapping and multi-scattering:} Studies have shown that IR trapping can be important in boosting the momentum of the gas to launch strong outflows from the galaxies \citep{bieri17,costa_2018a}. To estimate the extent of momentum boosting due to IR, we perform the below calculation in the disc plane for Sim.~B (perpendicular jet). We first calculate the density weighted mean velocity (for $n>1\,\mathrm{cm^{-3}}$) in the disc plane, which is, $v \approx $ 342 $\mathrm{km/s}$. This gives an upper limit of IR optical depth for efficient radiation coupling with the gas, which is, $\tau_{\mathrm{max}}=\frac{c}{3v}=350.54$, and an average estimate for the ratio of flow time and trapping time is: $ \frac{c}{3\tau_{\mathrm{max}}v}=13.12$ \citep{costa_2018a}.
We use a wavelength range of $7.8\times 10^{-4}\mathrm{mm}-1 \, \mathrm{mm}$ to estimate the IR optical depth, which gives a maximum value of 26.72. For this, the \textsc{Cloudy} computation was made to run until it achieves the default stopping criteria ($T \leq$ 4000 K), so that we can determine the optical depth well into the deeper regions. The average optical depth is 3.34.
This gives an upper estimate of momentum boost to be $\frac{\dot{p}}{L/c}=\tau_{\mathrm{max}}$ = 26.72, similar to the values observed in \citet{costa_2018a}, and an average boost of 3.34.

\section{Radiation vs Jet heating}
\label{rad-jet-heat}
\begin{figure}
\centering
\includegraphics[width=0.6\linewidth]{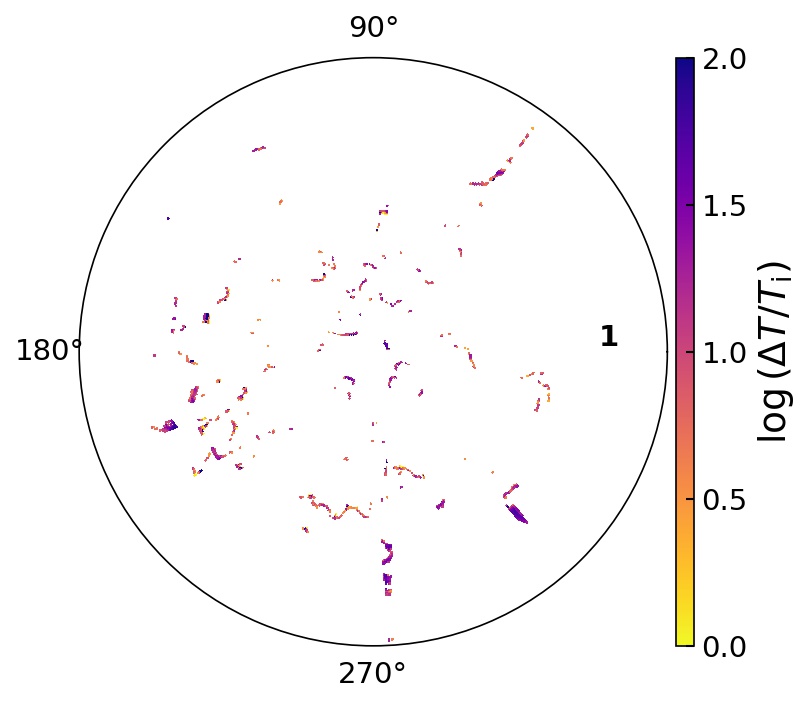}
\includegraphics[width=0.7\linewidth]{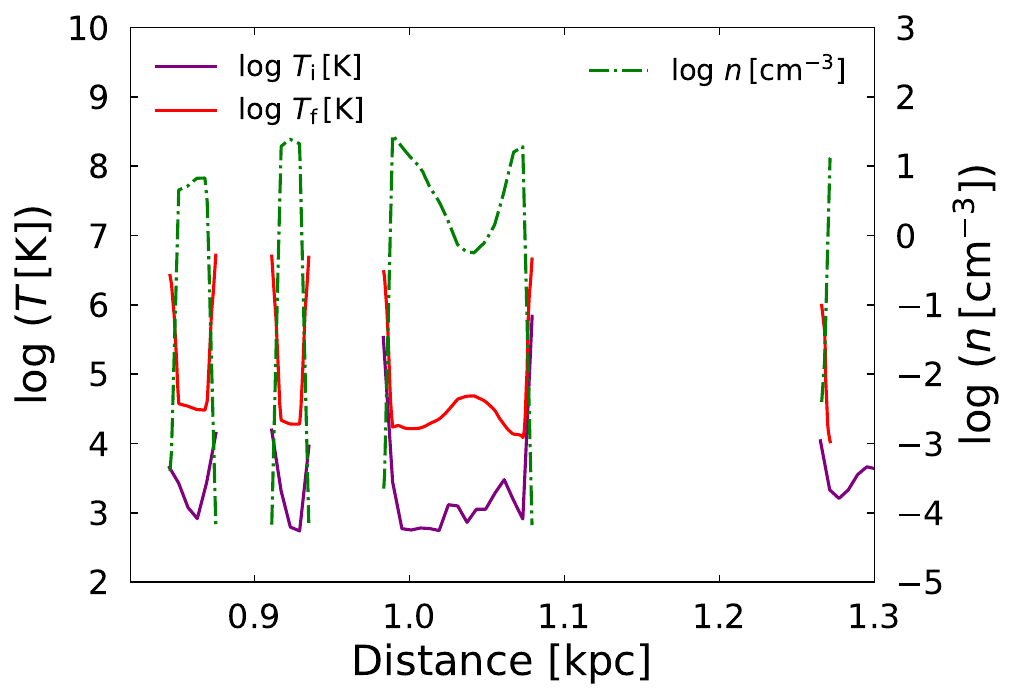}
\caption{Comparison of radiation vs jet heating in the disc plane of Sim.~B (perpendicular jet) at 0.68 Myr. \textbf{Top:} The ratio of change in temperature and initial temperatures (from simulation) for cells with density greater than $1\,\mathrm{cm^{-3}}$. The plot is shown up to a radius of $\sim1$~kpc (shown in bold digit). \textbf{Bottom:} An LOP in the disc plane (close to $\phi=270^\circ$) showing the input density (green), temperature before ($T_\mathrm{i}$ in purple) and after RT ($T_\mathrm{f}$ in red). The temperatures and density are shown to represent only the dense regions, and the collisionally ionized regions (see Sec.~\ref{shocks}) are removed.}
\label{fig:temp_comp}
\end{figure}

 In this section we investigate the effect of AGN radiation and jet heating by comparing the temperatures in the disc region of Sim.~B at 0.68 Myr, which is shown in Fig.~\ref{fig:temp_comp}. The top panel shows that some of the dense regions ($n>1~ \mathrm{cm^{-3}}$) in the disc plane are being heated to temperatures higher than the cooling floor of the simulation ($\sim 1000$ K). The bottom panel shows the one-dimensional plot for an LOP, indicating that AGN heating dominates and raise the temperature of some of the dense regions up to few orders of $10^4$ K, which are typical photoionization equilibrium temperatures \citep{dopita_2003}, and are observed for photoionized gas in several galaxies \citep{humphrey2008,nicole08}.

An estimate of the fraction of energy input ($\mathrm{erg\,cm^{-3}}$) due to radiation as compared to the jet for Sim.~B gives, 
\begin{equation}
    \frac{\sum n(T_f- T_i)}{\sum n T_i} = 4.2\times 10^{-3}
\end{equation}
where the sum in the numerator includes the regions where $T_f>T_i$ (temperature after RT is larger than the initial temperature) and a density cut-off of $1~\mathrm{cm^{-3}}$ is used.
These regions constitute only 1.28$\%$ of the total gas mass with ($n\geq1~\mathrm{cm^{-3}}$) in the disc (also see Table~\ref{tab:tab3}). 
\bsp	
\label{lastpage}
\end{document}